\documentstyle[12pt]{article} \long\def\del#1{ }         % Thu. Apr. 30, 1992
\let\3=\ss \catcode`\"=\active \let"=\"
\let\ssk=\smallskip  \let\bsk=\bigskip
  
   \def\ve{\vfil\eject}

\let\a=\alpha   \let\d=\delta 
  \let\th=\theta  
\let\l=\lambda    \let\p=\pi 
   \let\c=\chi
 
\let\Ph=\phi

\def\0{\over }    \def\1{\vec }   \def\2{{1\over2}} \def\3{{\ss}}
\def\4{{1\over4}} \def\5{\bar }   \def\6{\partial } \def\7#1{{#1}\llap{/}}
\def\8#1{{\textstyle{#1}}}        \def\9#1{{\bf {#1}}}
\def\_#1{$\underline{\hbox{#1}}$} \def\~#1{$\overline{\hbox{#1}}$}

\def\<{\langle } \def\>{\rangle }  
 
\def \({\left( } \def \){\right) }

  \let\ex=\times   \let\aus=\in
      \let\and=\wedge
     
\def\|#1{{}_{\bigg|_{#1}}}

\def\mao#1{\mathop{\rm {#1}}\nolimits}  \def\tr{\mao{tr}} 
  
\def\gcd{\mao{gcd}} 

\def\pmbf#1{\setbox0=\hbox{${#1}$}   \kern-.025em\copy0\kern-\wd0
      \kern.05em\copy0\kern-\wd0     \kern-.025em\raise.0433em\box0 }
   
   \def\ch{{\cal H}}
\def\cl{{\cal L}}  
\def\inbar{\vrule height1.5ex width.4pt depth0pt}
\def\IN{\relax{\rm I\kern-.18em N}}    \def\ZZ{\relax{\sf Z\kern-.4em Z}}
\def\IP{\relax{\rm I\kern-.18em P}}
\def\IQ{\relax\,\hbox{$\inbar\kern-.3em{\rm Q}$}}
\def\IR{\relax{\rm I\kern-.18em R}}
\def\IC{\relax\,\hbox{$\inbar\kern-.3em{\rm C}$}}

\def\beq{\begin{equation}} \def\eeq{\end{equation}} \def\eql#1{\label{#1}\eeq}
\def\bea{\begin{eqnarray}} \def\eea{\end{eqnarray}} 
\def\fnote#1#2{\begingroup\def\thefootnote{#1}\footnote{#2}
           \addtocounter{footnote}{-1}\endgroup}

\def\plb#1 #2 {Phys. Lett. {\bf B#1} #2 }
\def\phr#1 #2 {Phys. Rep. {\bf  #1} #2 } 
\def\npb#1 #2 {Nucl. Phys. {\bf B#1} #2 }
\def\aph#1 #2 {Ann. Phys. {\bf #1} #2 }  
\def\jmp#1 #2 {J. Math. Phys. {\bf #1} #2 }
\def\prd#1 #2 {Phys. Rev. {\bf D#1} #2 }
\def\prl#1 #2 {Phys. Rev. Lett. {\bf #1} #2 }
\def\rmp#1 #2 {Rev. Mod. Phys.  {\bf #1} #2 }
\def\zpc#1 #2 {Z. Phys. {\bf #1C} #2 }
\def\cmp#1 #2 {Comm. Math. Phys. {\bf #1} #2 }
\def\mpl#1 #2 {Mod. Phys. Lett. {\bf A#1} #2 }
\def\ijmp#1 #2 {Int. J. Mod. Phys. {\bf A#1} #2 }

\def\IP{\relax{\rm I\kern-.22em P}}
\def\ch{D}    \def\PP{Poincar\'e polynomial }
\def\tbf#1:{{\noindent\bf #1:}}  
\def\tt{\tilde\theta} \def\ngb{\bar n_g} \def\thi{\tilde\th_i}

\def\figuresonly{\pagestyle{empty}\figa\ve\figb\ve\figc\figd\ve
             \fige\end{document}}
\long\def\old#1\endold{{ }}                \def\oldansw{o } \def\cutansw{c }
\newcount{\npre} \npre=1  \def\negansw{s } \def\figansw{f } \def\textansw{t }
\def\ifpre{\ifnum\npre=1 } \def\ifsub{\ifnum\npre=0 }        \def\cut#1{#1}
\def\askversion{\message{
Preprint (p) / submit (s) / text only (t) / figures only (f):  (p/s/t/f)? }
    \read-1 to\answ \ifx\answ\negansw \npre=0 \else \npre=1 \fi
    \ifx\answ\figansw { } \else \def\figuresonly{ }   \fi
    \ifx\answ\oldansw \def\old##1\endold{{\small ##1}}\fi
    \ifx\answ\textansw \npre=2 \else \message{
Cut figures (use 'c' in case of memory problem):  (c/n)? }
    \read-1 to\answ\ifx\answ\cutansw \def\cut##1{}\npre=7\fi\fi \figuresonly }

\def\bpic{\begin{picture}} \def\epic{\end{picture}} \thicklines
\def\lab#1)#2#3{\put#1){\makebox(0,0)[#2]{\small #3}}}
\def\putlin#1,#2,#3,#4,#5){\put#1,#2){\line(#3,#4){#5}}} %\putlin(x,y,dx,dy,l)
\def\putvec#1,#2,#3,#4,#5){\put#1,#2){\vector(#3,#4){#5}}}

\unitlength=0.15truemm  \newcount{\vdiv}    % divide vert. length by \vdiv
\newcount{\wi} \newcount{\he}              % \hq \vq ... h./vert. quantities
\newcount{\auxc}\newcount{\ptsize}
\def\pt(#1,#2){\auxc=#1 \divide\auxc by\vdiv \put(#2,\auxc){\circle*{\ptsize}}}
\def\.#1,#2){\auxc=#1 \divide\auxc by\vdiv \put(#2,\auxc){\circle*{\ptsize}}}
\def\hlab#1{\putlin(#1,5,0,-1,10)\lab(#1,-10)t{#1} }
\def\vlab#1{\auxc=#1\divide\auxc by\vdiv
          \putlin(5,\auxc,-1,0,10)\lab(-10,\auxc)r{#1} }
\def\hopl#1{\auxc=\he \divide\auxc by\vdiv
\vbox{\begin{center} \bpic(\wi,\auxc){#1}
\putvec(0,0,1,0,\wi)  \lab(\wi,-10)t{\hq}
\putvec(0,0,0,1,\auxc) \lab(-10,\auxc)r{\vq}
\hlab{100} \hlab{200} \hlab{300} \hlab{400}
\vertlab
\epic \\[1cm] \capt \end{center}}}

\def\figa{\he=3500 \vdiv=4  \ptsize=2 \def\vq{$d$} \def\hq{$\bar n_g$}
\def\capt{Fig. 1: A plot of $d$ vs. the number of $\bar{27}$
	representations for 4 variables.}
\def\vertlab{\vlab{500} \vlab{1000} \vlab{1500}\vlab{2000}\vlab{2500}
	      \vlab{3000}  }
\hopl{{\input d-ngb-4 }}       }

\def\figb{\he=2000 \vdiv=4  \ptsize=2 \def\vq{$d$} \def\hq{$\bar n_g$}
\def\capt{Fig. 2: A plot of $d$ vs. the number of $\bar{27}$
	representations for 5 variables.}
\def\vertlab{\vlab{500} \vlab{1000} \vlab{1500}}
\hopl{{\input d-ngb-5 }}       }

\def\figc{\he=2000 \vdiv=4  \ptsize=2 \def\vq{$d$} \def\hq{$\bar n_g$}
\def\capt{Fig. 3: A plot of $d$ vs. the number of $\bar{27}$
	representations for 6 variables.}
\def\vertlab{\vlab{500} \vlab{1000} \vlab{1500}}
\hopl{{\input d-ngb-6 }}       }

\def\figd{\he=2000 \vdiv=4  \ptsize=2 \def\vq{$d$} \def\hq{$\bar n_g$}
\def\capt{Fig. 4: A plot of $d$ vs. the number of $\bar{27}$
      representations for 7 variables.}
\def\vertlab{\vlab{500} \vlab{1000} \vlab{1500}}
\hopl{{\input d-ngb-7 }}       }

\def\fige{\def\pt(##1,##2){\put(##2,##1){\circle*{2}}}
\he=500 \vdiv=1  \def\vq{$p$} \def\hq{$\bar n_g$}
\def\capt{Fig. 5: A plot of $p=\phi(d)/2-1$ vs. the number of $\bar{27}$
	representations for all non-degenerate configurations.}
\def\vertlab{\vlab{100} \vlab{200} \vlab{300} \vlab{400}}
\hopl{{\input p-ngb }}      }

\begin{document} \askversion  {\hfill CERN-TH.6461/92}\vskip-9pt
                              {\hfill TUW--92--06}
\vskip 15mm
\begin{center}{\large NO MIRROR SYMMETRY IN LANDAU-GINZBURG SPECTRA!}
\vskip 12mm
       Maximilian KREUZER\fnote{\star}{e-mail: kreuzer@cernvm.cern.ch}
\vskip 5mm
      Theory Division, CERN\\
       CH--1211 Geneva 23, SWITZERLAND
\vskip 7mm               and
\vskip 7mm        Harald SKARKE\fnote{\#}{e-mail: hskarke@email.tuwien.ac.at}
\vskip 5mm
      Institut f"ur Theoretische Physik, Technische Universit"at Wien\\
      Wiedner Hauptstrasse 8--10, A--1040 Wien, AUSTRIA

\vfil                        {\bf ABSTRACT}                \end{center} \ssk

   We use a recent classification of non-degenerate quasihomogeneous
polynomials to construct all Landau-Ginzburg (LG) potentials for N=2
superconformal field theories with c=9 and calculate the corresponding
Hodge numbers. Surprisingly, the resulting spectra are less symmetric than
the existing incomplete results.
It turns out that models belonging to the large class for which an explicit
construction of a mirror model as an orbifold is known show remarkable
mirror symmetry. On the other hand, half of the remaining 15\% of all models
have no mirror partners.
This lack of mirror symmetry may point beyond the class of LG-orbifolds.
\vfil \noindent
CERN-TH.6461/92\\
TUW--92--06\\
April 1992
\thispagestyle{empty} \eject \setcounter{page}{1} \pagestyle{plain}
\ifsub \baselineskip=20pt \else \baselineskip=15pt \fi

\section{Introduction}

One of the most promising strategies for obtaining realistic physical
models from the Heterotic string is the consideration of N=2 superconformal
field theories (SCFT) with integer left U(1) charges for a description
of the internal degrees of freedom, which lead to N=1 space-time
supersymmetric models \cite{N2}. A large class of N=2 SCFTs with equal
rational left and right U(1) charges can be obtained by means of a
Landau-Ginzburg (LG) description \cite{mvw,lvw}. N=2 world-sheet
superconformal invariance is assumed to imply
non-renormalization of the superpotential $W$ in the action
\beq
\cl=\int d^2zd^4\th K(\Ph_i,\bar\Ph_i)+\(\int d^2zd^2\th W(\Ph_i)+c.c.\) \eeq
Thus, with a quasi-homogeneous potential $W(\l^{n_i}\Ph_i)=\l^dW(\Ph_i)$,
this action should describe a conformal model at the renormalization group
fixed point \cite{z}.
Orbifolding these theories by discrete groups containing the canonical
$\ZZ_d$ symmetry of the potential \cite{v,iv} leads to models with the
desired property of integral left U(1) charges. Canonically orbifolded
LG-theories with 5 variables in the potential are directly related to
Calabi-Yau (CY) manifolds described as the zero-locus of the equation
$W(z_i)=0$ in weighted projective space \cite{gvw,cls}.

\old
Landau-Ginzburg (LG) potentials provide a very general class of heterotic
string vacua \cite{mvw,lvw}. N=2 world-sheet superconformal invariance, which
is necessary for N=1 space-time supersymmetry \cite{N2}, is assumed to imply
non-renormalization of the superpotential $W$ in the action
\beq
\cl=\int d^2zd^4\th K(\Ph_i,\bar\Ph_i)+\(\int d^2zd^2\th W(\Ph_i)+c.c.\) \eeq
Thus, with a quasi-homogeneous potential
\beq W(\l^{n_i}\Ph_i)=\l^dW(\Ph_i),  \eeq
this action should describe a conformal model at the renormalization group
fixed point \cite{z}.

LG theories are closely related to Calabi-Yau manifolds,
which can be constructed as the zero-locus of the equation
$W(z_i)=0$ in weighted projective space \cite{gvw,cls}.
Promising phenomenology, on the other hand, presumably requires
orbifoldization of the LG-models \cite{v,iv}.
By including discrete torsion \cite{iv},
one even arrives at the much larger class of (2,0)-models, which in general
have no more (immediate) geometric interpretation.
\endold

An interesting result of the calculation of a large class of Calabi-Yau
manifolds in weighted $\IP_4$ by Candelas et al. \cite{cls}
was the observation of an approximate symmetry of the spectra under the
exchange of the Hodge numbers $b_{21}$ and $b_{11}$, viz. the
$27$ and $\bar{27}$ representations of $E_6$. This so called mirror symmetry
(MS), which was predicted from CFT-arguments by Dixon and Gepner,
%{\bf citation! }
is a powerful computational tool, because only one set of Yukawa couplings
is subject to (non-perturbative) quantum corrections. MS may thus be used
to calculate non-perturbatively all Yukawa couplings for a Calabi-Yau
manifold once the mirror partner is known~\cite{co}. Partial results on
abelian orbifolds of LG-models \cite{kss} showed a further increase
of MS (at least in the naive counting of just comparing spectra)
to 94\%, in accordance with expectations based on recently found
techniques for explicit constructions of mirror models \cite{gp,bh}.

In the present paper we report on a complete computation of all
LG-potentials
with $c=9$ based on our recent classification of quasi-homogeneous polynomials
with non-degenerate critical points \cite{ks}.
Even if the resulting models are not too promising phenomenologically,
our calculations are a valuable step toward more realistic models since they
set the stage for a systematic investigation of orbifolds.
Our most striking result, however, is an actual decrease of symmetry in the
resulting spectra, which we will discuss in some detail below.

In section 2 we recall the essential results on the classification of
non-degenerate quasihomogeneous polynomials \cite{ks} and describe our
methods of calculation.
In section 3 we recall how to calculate the Hodge numbers, deriving some
useful formulas. We also discuss the conditions for ``factorization'' of a
LG-orbifold, which would lift the zeros of the Hodge diamond.
In sections 4 and 5 we present and discuss our results.

\section{Calculation of c=9 LG-potentials}

\subsection{Classification of non-degenerate quasihomogeneous polynomials}

In the following we summarize recent results \cite{ks} on the classification
of non-degenerate quasihomogeneous polynomials \cite{agv} on which our
calculations are based. Some important points are illustrated by examples.

A polynomial $W(X_i)$ is said to be quasihomogeneous of degree
$d$, if there exist integers $n_i$ such that
$W(\l^{n_i}X_i)=\l^dW(X_i)$. $q_i=n_i/d$ is called the weight of
$X_i$. We call the space of all polynomials with a given
weight structure $(n_i,d)$ a configuration.
If $W$ has an {\it isolated} critical point at $X_i=0$ it is called
non-degenerate. A configuration is called non-degenerate if it contains a
non-degenerate member.

The local algebra of $W$ is defined as the ring of all
polynomials in the $X_i$ modulo the ideal generated by the gradients
$\6W/\6X_i$.
It is finite dimensional if and only if $W$ is non-degenerate.
In this case the highest weight occurring in the local algebra is
given by the singularity index $\ch=\sum_i(1-2q_i)$. The central charge
of the N=2 superconformal field theory whose Landau-Ginzburg potential
is $W$ is given by $c=3\ch$~\cite{lvw}. The Poincar\`e
polynomial $P(t)$ is defined as the generating function for the number of
basis monomials of the local algebra of a specific degree of quasihomogeneity,
i.e. the number of states of a given conformal weight. It can be computed
with the formula
\beq P(t)=\prod{1-t^{d-n_i}\0 1-t^{n_i}}. \eql{pp}
%Note that this expression is not a polynomial in
%$t$ but rather in $t^{1/d}$.
%For convenience, however, we will refer to $P(t)$ and not to $P(t^d)$ as the
%Poincar\`e polynomial.
Of course, a necessary condition for non-degeneracy
is given by the requirement that the r.h.s. of this equation be a
polynomial. We will call such a configuration {\it almost non-degenerate}.
The r.h.s. of eq. \ref{pp} will be a polynomial if and only if all zeros
of the denominator (counted with their multiplicities) are also zeros
of the numerator. This means that the set of all multiples of the
numbers $1/(d-n_i)$ contains the set of the multiples of
the $1/n_i$. This is equivalent to the statement that the set of divisors
of the $(d-n_i)$'s contains the set of divisors of the $n_i$,
(again, multiplicities are to be taken into account).

%We will now briefly summarise those
The main result of ref. \cite{ks} uses the following definitions, which
suggest a graphical description of the structure of a quasihomogeneous
polynomial.
A variable $X$ is called a root if the polynomial $W$ contains a term
$X^a$. A monomial $Y^aZ$ is called a pointer at $Z$.
$a$ is called the exponent of $X$ or $Y$, respectively.
We recursively define
a link between two expressions, which may themselves be variables or links,
as a monomial depending only on the variables occurring in these expressions.
A link may further be linear in an additional variable $Z$,
which does not count as a variable of the link. In this case we
say that the link points at $Z$, thus extending the previous definition
of a pointer. Of course a specific monomial occurring in $W$ can have
more than one interpretation as a link or pointer.
Given $W$, we call any graph (not necessarily the maximal one) whose lines
allow the above interpretation in terms of monomials in $W$ a graphic
representation of $W$. We represent variables by dots, pointers
by arrows and links by dashed lines. A graph without links
will be called ``skeleton graph''.

\tbf Theorem 1: For a configuration a necessary and sufficient condition for
non-degeneracy is that it has a member which
can be represented by a graph with:\\
1. Each variable is either a root or points at another variable.\\
2. For any pair of variables and/or links pointing at the same variable Z
   there is a link joining the two pointers
   and not pointing at $Z$ or any of the targets of the sublinks
   which are joined.

\tbf Example 1: To illustrate these conditions we consider the configuration
$\{n_i\}=\{5$, 6, 7, 8, 9, 11, $12\}$ and $d=29$.
In order that $X_i$ can point at $X_j$ $n_i$ has to divide the
$j^{th}$ entry in the list $\{d-n_i\}=\{24,23,22,21,20,18,17\}$. Thus we have,
in an obvious notation, $1\to5\to 6\to 3\to 4\to 1$ and  $7\to1$. For
$X_2$ there are 2 possibilities: it may point at $X_1$ or at $X_6$. We choose
the first one and arrive at the polynomial
\beq W_{\rm skeleton}=X_1^4X_5+X_5^2X_6+X_6^2X_3+X_3^3X_4+X_4^3X_1 +X_2^4X_1+
       X_7^2X_1.                                \eeq
We now have a triple pointer at $X_1$ and thus need the 3 links
$(24)[1]$, $(47)[1]$ and $(27)[1]$, where numbers in parenthesis
represent the variables of a link, whereas square brackets indicate the
forbidden targets. $X_j$ can appear as a target of the first
link only if $d-n_j$ is a multiple of $\gcd(n_2,n_4)=2$, thus the first link
can point at $X_3$, $X_5$ or $X_6$ and we again choose the first possibility.
The targets of the other links are unique (at this stage all links have to
be pointers because $d$ is prime and none of the $\gcd$'s of the involved $n_i$
is 1). We choose to add the following monomials to the potential,
\beq W_{\rm links}=X_3(X_2X_4^2) + X_5(X_4X_7) + X_6(X_2^2),               \eeq
which in return generate 3 new double pointers and thus imply the respective
links $(246)[13]$, $(471)[15]$ and $(275)[16]$. The first 2 of these have
$\gcd(n_i,n_j,n_k)=1$ and can, e.g., represent the monomials
$X_2^2X_6$ and  $X_7^2X_1$, which are already present in the potential.
The last link has to point at $X_4$, which can represent the second monomial
in $W_1$. Again, the resulting double pointer finally implies the link
$(2753)[164]$ with $\gcd(6,12,9,7)=1$. This requires an additional
contribution to the potential, which can now be completed to the
non-degenerate polynomial
\beq W=W_{\rm skeleton}+W_{\rm links}+X_2X_3^2X_5. \eeq
Our fairly complicated example illustrates some important points:
A single monomial can correspond to several different links and neither
the choice of a link nor the choice of the corresponding monomial is unique.
This does not matter for checking the non-degeneracy criterion.
Different choices, however, may of course lead to different discrete
symmetries.

We have also seen that divisibility conditions play an important role for
the criticality properties of a configuration. In fact, there is a close
relation to the Poincar\'e polynomial \cite{ks}:\\
\tbf Lemma 1: The necessary condition for non-degeneracy that
the expression (\ref{pp}) for the Poincar\'e polynomial is a polynomial
is equivalent to the criterion of theorem 1 if one omits the requirement
that all exponents in the link monomials have to be non-negative.

\tbf Example 2:  To illustrate this connection we consider the
configuration $\{n_i\}=\{1,1,6,14,21\}$, $d=43$ for which the expression
(\ref{pp}) is a polynomial. As above we choose the ``skeleton'' polynomial
\beq  W_{\rm skeleton}=X_1^{42}+X_2^{42}+X_1X_3^7+X_1X_4^3+X_2X_5^2 \eeq
which has only one double pointer, requiring the link $(34)[1]$.
As $\gcd(6,14)=2$ this link can only point at $X_2$ or at $X_5$.
In the second case we would need a monomial $X_3^aX_4^bX_5$. Consistency with
quasihomogeneity implies $6a+14b=22$ which, however, has no solution with
both $a$ and $b$ non-negative (by choosing $X_2$ as the target we would
be able to find a link, but would be stuck in the same way one step later).
Thus we see that this configuration is degenerate, although the divisibility
conditions are fulfilled. There are 452 such  ``almost non-degenerate''
configurations with a formal $\ch$ of 3, most of them have 5 variables.

We now know how to check whether a given configuration is degenerate. In order
to be able to calculate all non-degenerate configuations with $c=9$ we
still need to restrict the possible (skeleton) graphs and exponents to a
finite set. We call variables trivial if they correspond to terms $X^2$.
Trivial variables have weights $q=1/2$ and therefore do not contribute to
$\ch$, nor to the local algebra, as they can be eliminated by $\6W/\6X=0$.
The same is true for Lagrange multipliers (i.e. variables which appear only
linearly) in non-degenerate configurations, which may have
any weight $q$ and can be eliminated pairwise together
with their ``target'' field with weight $1-q$. All other fields have
weight $q>1/2$. The following two lemmata complete our review of the results
we need from {\it ref.}~\cite{ks}.

\tbf Lemma 2: For every non-degenerate configuration $\ch$ is
greater than or equal to 1/3 times the number of non-trivial variables.
For $\ch=3$ this means that the number of non-trivial variables is between 4
and  9. In the case of 9 variables the only configuration is
$n_i=1$ and $d=3$.

\tbf Lemma 3: Given a positive rational number $\ch$, there is only a
finite number of non-degenerate configurations whose index is $\ch$.
For a non-degenerate quasihomogeneous polynomial
with $\ch=3$ the number of exponents $a_i>18$ or $a_i>84$
is smaller than 3 or 2, respectively.

\subsection{The calculations}

Using these results we have calculated all solutions to $\ch=3$ in 3 steps.
First we have calculated all inequivalent skeleton graphs which can
yield such configurations. As the solution for $N=9$ variables is unique,
we need graphs with between 4 and 8 points, the numbers of which are
19, 47, 130, 343, and 951, respectively. We thus had to invstigate 1490
skeleton graphs, of which 175 have no double pointers and thus do not
require any links. Remarkably, this comparatively small number of graphs
already generates the vast majority of configurations (i.e. 9108 out of 10839).

It would of course be virtually impossible to explicitly construct all
the required links for the remaining graphs. Fortunately, however, there
is only a finite set of possibilities for the exponents in the skeleton
graphs when one requires non-degeneracy, and these exponents already
determine the configuration.
Lemma 3 restricts all
but one of the exponents to be less than 85. One may thus first go
through the (inequivalent) choices of a variable for the free exponent $a$ and
use the bounds on all other exponents \cite{ks} to generate all possible cases
and calculate $a$ from the condition $\ch=3$.
If $a$ turns out integer one can check whether the expression (\ref{pp})
is a polynomial and insert the resulting almost non-degenerate
configurations into an ordered list.  There is a
complication to this procedure by the fact that in some cases
$\ch=3$ independent of the last exponent. An extreme example of this kind
is the polynomial
\beq W=X_1^aX_2+X_2^bX_3+X_3^6X_1+ X_1 X_4^2+X_1 X_5^3+X_2X_6^2+X_2X_7^2 \eeq
for which this is even the case for arbitrary $a$ and $b$.
A careful examination by
hand shows that in these cases the open exponent cannot be
greater than 84 without violating the link criterion.

In a third step we have checked non-degeneracy using theorem 1 and calculated
the Hodge numbers (see below).
As the calculation of skeleton graphs and the criterion for non-degeneracy
are recursive it was straightforward to implement the sketched procedure.
All of our programs are written in the language C. Of course one has to
be careful to avoid integer overflow. With the variable type ``long int''
we were, however, safe by a factor of 100, which we derived both
theoretically (from limits on the exponents) and ``experimentally''.
A complete run of our programs takes more than a week on an HP 9000/720
workstation.

In order to check our results we have also implemented an independet
algorithm for finding all configurations.
Once the limits on  $d$ are known a simple program can generate all
partitions of $d(N-3)/2$. For these one can check whether the $n_i$ have
no common divisor and whether (\ref{pp}) is a polynomial.
If an upper bound for $d$ is known one can thus generate the list
of all almost non-degenerate configurations.
Unfortunately the number of possibilities grows roughly like $d^N/N!$.
For small $d$, however, this is a very simple and efficient procedure.
One can work on the efficiency for large $d$ by requiring from
the start that the partition admits a pointer structure. With such an improved
algorithm we have checked our results up to $d$ equal to $5000$, $2000$,
$700$, $250$ and $200$ for $N=4,\ldots,8$. This has to be compared to the
maximal values of $d$ in the direct computation which are 3486, 1743, 1806,
600 and 384 for the respective numbers of variables.

\section{Calculation of Hodge numbers}

In order to use a LG-model for constucting a consistent Heterotic string we
need to project onto integral (left) charges \cite{gso,v}.
The resulting chiral ring determines the gauge multiplets: The chiral primary
fields with $(q_L,q_R)=(1,1)$ yield 27's of $E_6$, whereas the states with
charges (1,2), which are related to the (1,-1) states by spectral flow,
end up as anti-generations.  In the following
we will only consider the simplest case of the canonical $\ZZ_d$ orbifold.
In case of 5 variables its chiral
ring is related to the cohomology ring of the corresponding Calabi-Yau
manifold in the weighted projective space $\IP_4$ \cite{gvw}.
We will call the respective dimensions Hodge numbers
also in the present LG context.

The formulas for calculating these numbers
were given by Vafa \cite{v}. The numbers $p_{ij}$ of states with
$(q_L,q_R)=(i,j)$ are the coefficients of $t^i\bar t^j$ in the generalized \PP
\fnote{1}{When comparing
with the corresponding Calabi-Yau one shoud note that $p_{11}$ corresponds to
$b_{12}$ and vice versa.}
\beq P(t,\5t)=\tr\,t^{J_0}\5t^{\5J_0}= Q(t^{1\0d},\5t^{1\0d})|_{int},\eql{gpp}
with
\beq Q(t^{1\0d},\5t^{1\0d})=\sum_{0\le l<d}\prod_{\tt_i\aus\ZZ}
    {1-(t\5t)^{1-q_i}\01-(t\5t)^{q_i}}\prod_{\tt_i\not\aus\ZZ}(t\5t)^{\2-q_i}
	   \({t\0\5t}\)^{\tt_i-\2},                                \eql{uf}
where  $\th_i=lq_i$ and $\tt_i=\th_i-[\th_i]$ is the non-integer part of
$lq_i$. The subscript $int$ means that only integral powers of $t$ and
$\bar t$ are kept in the expression for $P$.

It is well known that for Calabi-Yau manifolds $b_{01}=b_{02}=0$ at least for
$\chi\ne0$, where $\chi$ is the Euler number $\chi=2(b_{11}-b_{12})$ \cite{c}.
In the context of LG models formula (\ref{uf}) implies that the exponent of
$t$ can only vanish if $\sum_{\tt_i\not\aus\ZZ}(\tt_i-q_i)=0$. For $\th_i=q_i$
this gives $p_{03}=1$. For the other $p_{0j}$ we have\\
\tbf Theorem 2: Contributions to $p_{01}$ and $p_{02}$ in (\ref{gpp}) can only
arise if there is a subset of the $q_i$ with $\sum_{\rm subset}(1-2q_i)\in
\ZZ$ and an element of the $\ZZ_d$ which acts on this subset like the
generator of the $\ZZ_d$ and does not act on the other fields.
In this case the
$\ZZ_d$ orbifold factorizes into a product of canonical $\ZZ_{d_i}$
orbifolds of LG-models with integer indices $D_i$ with $\gcd(d_i,d_j)=1$ and
$\sum D_i=D$. For $D=3$ this implies vanishing Euler number because one of the
factors must have $D=1$.\\
{\it Proof}: If $X_i$ occurs in the potential
as $X_i^\a$, $\tilde\th_i$ will be of the form $\l/\a$ and therefore larger
than or equal to $q_i=1/\a$. If $X_i$ occurs as $X_jX_i^\a$, we consider
$\d_i=\tilde\th_i-q_i$. Obviously if $\d_i< 0$ then $\d_j=1-\a\tilde\th_i
-(1-\a q_i)=-\a\d_i$,
yielding $\d_i+\d_j=-(\a-1)\d_i>0$. There is a slight subtlety concerning the
possibility that several different $X_i$ with $\d_i<0$ might belong to the
same $X_j$. If we consider all $X_i$ with $\d_i$ smaller than some given
negative value $\d$, calculate $dW$ and then set all other $X_k$ to zero,
non-degeneracy implies that we must have at least as many equations left as
there are $X_i$. These equations must come from $dW/dX_j$ with $\d_j\ge-2\d$.
Therefore, if we introduce a ranking of the $X_i$ with respect to the
absolute value of $\d_i$ and a corresponding ranking of the $X_j$, we see that
$\d_i+\d_j$ will always be positive for corresponding $i$'s and $j$'s,
thereby proving the first statement of the theorem.
%This result allows us to derive an analogous statement in the LG-context.
%\tbf Theorem 2: The coefficients of $t$ and $\bar t$ in (\ref{gpp}) can be
%non-zero only if the condition of lemma 4 is fulfilled.  \\
Now consider an integer $l$ in (\ref{uf}) and an appropriate
ordering of the variables such that $\thi=q_i$ for $0<i\le I$ and $\thi=0$ for
$I<i\le N$. We define $D_1=\sum_{i\le I}(1-2q_i)$ and represent $q_i$ by
simplified fractions $q_i=r_i/s_i$. %with $\gcd(r_i,s_i)=1$.
This implies $\gcd(s_i,s_j)=1$
for $i\le I<j$, as $s_i$ divides $l-1$ and $s_j$ divides $l$. With $d_1$ and
$d_2$ being the least common multiples of the $s_i$ with $i\le I$ and $i>I$,
respectively, $\ZZ_d=\ZZ_{d_1}\ex\ZZ_{d_2}$ and the complete expression
(\ref{gpp}) factorizes into a product of canonical LG-orbifolds with integer
indices. //.

In the following we assume $p_{01}=0$ (we will reconsider factorization at
the end of  section 4), thus
\beq P(t,\bar t)=(1+t^3)(1+\5t^3)+n_g(t\5t+t^2\5t^2)+\5n_g(t\5t^2+t^2\5t).
                                       \eql{Ptt}
Vafa has also given a formula for the Euler number which is much simpler
than the ones above for the Hodge numbers. This formula and an
analogous one for the sum of all Hodge numbers, which in our context
is equal to $4+2n_g+2\bar n_g$, can be derived in the following way:
The projection of $Q$, which is a polynomial in $t^{1\0 d}$ and
$\5t^{1\0 d}$, onto $P$, can be achieved through
\beq P(t, \5t)=d^{-2}\sum_{j=1}^d\sum_{k=1}^dQ(e^{2\p i{j\0d}}t^{1\0d},
     e^{2\p i{k\0d}}\5t^{1\0d}).                                       \eeq
With $P(1,1)=4+2n_g+2\5n_g$ and $P(-1,-1)=2(n_g-\5n_g)$ we get
\beq 4+2n_g+2\5n_g=d^{-2}\sum_{j=1}^d\sum_{k=1}^dQ(e^{2\p i{j\0d}},
     e^{2\p i{k\0d}})                                                  \eeq
and
\beq 2n_g-2\5n_g=d^{-2}\sum_{j=1}^d\sum_{k=1}^dQ(e^{2\p i{2j+1\02d}},
     e^{2\p i{2k+1\02d}}).                                              \eeq
The calculation of these expressions is straightforward, using
$\sum (1-2q_i)=3$ and
\beq {1-e^{2\p i\,j(1-q_i)}\0 1-e^{2\p i\,jq_i}}={1-q_i\0q_i}\,{\rm or}\,
     e^{-2\p i\,jq_i}                                                    \eeq
for $jq_i \in \ZZ$ or $jq_i \not\in \ZZ$, respectively. We find
(with the symbol ``$\cap$'' for the greatest common divisor of two integers)
\beq 4+2n_g+2\5n_g=d^{-1}\sum_{k=1}^d\sum_{l=1}^d (-1)^{(N-1)k(l-1)+N_l-
     N_{l\cap k}}\prod_{(l\cap k)q_i\in \ZZ}{1-q_i\0q_i}            \eql{sum}
and
\beq 2n_g-2\5n_g=d^{-1}\sum_{k=1}^d\sum_{l=1}^d (-1)^{(N-1)(kl-k-l)+N-
     N_{l\cap k}}\prod_{(l\cap k)q_i\in \ZZ}{1-q_i\0q_i},           \eql{vaf}
where $N$ denotes the total number of fields and $N_l$ denotes the number
of fields for which $lq\in\ZZ$.
By adding a trivial term $X^2$ to our
potential, we can always make $N$ odd\fnote{2}{For the $\ZZ_d$-orbifolds
this cannot lead to a doubling of the ground state because
$d=(2\sum n_i)/(N-3)$ has to be even for even $N$.},
thereby getting rid of the first expression in the exponent of $(-1)$.
We find that for odd $N$
\beq \chi=
    d^{-1}\sum_{k=1}^d\sum_{l=1}^d\prod_{(l\cap k)q_i\in \ZZ}{q_i-1\0q_i},\eeq
\beq 4+2n_g+2\5n_g=d^{-1}\sum_{k=1}^d(-1)^{N_k}\sum_{l=1}^d
	      \prod_{(l\cap k)q_i\in \ZZ}{q_i-1\0q_i}.           \eql{xbar}
One should note that these formulas allow calculations of $n_g$
and $\bar n_g$ in a much easier way.
It is quite remarkable that a generalisation of these formulas,
which make no reference to the actual structure of the local
algebra / chiral ring, seems to work for arbitrary orbifolds.

A useful estimate for the number of antigenerations is obtained with the
following argument: If $g$ is the generator of the canonical group and
$a\cap d=1$,
the action of $g^a$ will leave no $X_i$ invariant. Therefore the vacuum in the
sector twisted by such an element will be invariant under the action of any
group element. Each $g^a$ will add a term with coefficient 1
to the Poincar\'e polynomial. The elements corresponding to $a=1$ and $a=d-1$
give $\bar t^3$ and $t^3$, respectively, while all others contribute to
$\bar n_g$.
The number $\bar n_g$ of antigenerations therefore obeys $\bar n_g \ge
\phi(d)/2-1$, where $\phi$
is Euler's function, i.e. $\phi(d)$ is the number of integers $a$ with
$0<a<d$ and $a\cap d=1$.

\section{Results}

Table I lists the numbers of different types of potentials that we
have found.
The numbers on top of the columns indicate the numbers of
variables. The columns 1g, 2g and 3g show the numbers of 1-, 2- and
3-generation models, respectively. The
column c3 shows the number of models with Euler numbers that are odd
multiples of 6 (these are probably the best candidates for 3-generation
orbifold models).

\noindent
\begin{tabular}{||l|rrrrrr|rrrr|r||} \hline\hline
& 4 & 5 & 6 & 7 & 8 & 9 & 1g & 2g & 3g & c3 & Total      \\ \hline
Non-degenerate &2390&5165&2567&669&47&1&1&26&40&496&10839\\
Invertible&2069&4191&2239&568&40&1&0&5&20&286&9108\\
Not Invertible&321&974&328&101&7&0&1&21&20&210&1731\\
Degenerate&14&418&3&17&0&0&0&1&4&125&452\\
Almost Non-deg.&2404&5583&2570&686&47&1&1&27&44&621&11291\\
\hline\hline \end{tabular}\hfill\\[3mm]
\hfil Table I: Numbers of various types of models.
\old In lines 2 and 3 we list
the models which can/cannot be represented without links. In the last line
we count all configurations fulfilling the \PP condition.  \endold
\bsk

By invertible we mean those configurations which contain polynomials which
do not require links for non-degeneracy (i.e., they have as many monomials
as variables). These are precisely the models where each point in their
graphical representation is hit by not more than one pointer, i.e. where
an inversion of the direction of the arrows in the graph yields again a
sensible graph. According to Berglund and H"ubsch \cite{bh}, these models
have mirrors that can be represented as orbifolds of the model that one
obtains by inverting the directions of the arrows.
By degenerate we mean the almost non-degenerate configurations generated
by our program which violate the link criterion. It is remarkable
that these models yield integer Hodge numbers when one naively applies
formulas \ref{sum} and \ref{vaf}.

In figs. 1 -- 4 we have plotted $d$ versus $\ngb$ for the non-degenerate
models with 4, 5, 6 or 7 variables, respectively.
There is an apparent quantization of directions in the $d-\ngb$ plane,
with different ranges covered for $N$ even and for $N$ odd.
In fig. 5 we have plotted
$\phi(d)/2-1$ over $\ngb$ for all non-degenerate models.
The plot shows excellent agreement with the theoretical estimate of section 2.
A plot of $n_g+\ngb$ versus $\c$ has, of course, the symmetric form familiar
from refs. \cite{cls} and \cite{kss} and is not given here.

The numbers of spectra for the different types of models are: %There are
2997 different spectra from 10839 non-degenerate models,
2339 different spectra from 9108 invertible models and
3371 different spectra from 11291 almost non-degenerate models.
77\% of all spectra coming from non-degenerate, 92\% of all spectra coming
from invertible and only 69\% of all spectra coming from our almost
non-degenerate models have mirrors in their respective lists of spectra.
If we do not count spectra but configurations, this asymmetry is even more
striking:
In our set of non-degenerate configurations, only 216 out of 9107
invertible models (that's approximately 2\%) have no mirror partners,
whereas 856 out of 1731 non-invertible models, i.e. quite exactly half of them,
are singles.

The non-degenerate models gave rise to 342, the invertible models to 268
and all almost non-degenerate models to 411 different Euler numbers.
In our list of non-degenerate configurations we did not find
one Euler number (namely 22) contained in the list in \cite{cls}.
The 75 Euler numbers which are not contained there are shown in table II.

{}~\\
\begin{tabular}{|r|}
\hline
-506\\-450\\-416\\-390\\-364\\
\hline\end{tabular}\hfil
\hbox{\begin{tabular}{|r|}\hline
-356\\-344\\-320\\-294\\-286\\
\hline\end{tabular}}\hfil
\hbox{\begin{tabular}{|r|}\hline
-284\\-258\\-244\\-202\\-194\\
\hline\end{tabular}}\hfil
\hbox{\begin{tabular}{|r|}\hline
-190\\-172\\-166\\-162\\-142\\
\hline\end{tabular}}\hfil
\hbox{\begin{tabular}{|r|}\hline
-134\\-130\\-110\\-106\\-98\\
\hline\end{tabular}}\hfil
\hbox{\begin{tabular}{|r|}\hline
-94\\-74\\-68\\-62\\-58\\
\hline\end{tabular}}\hfil
\hbox{\begin{tabular}{|r|}\hline
-46\\-38\\-26\\-14\\-10\\
\hline\end{tabular}}\hfil
\hbox{\begin{tabular}{|r|}\hline
0\\2\\10\\34\\52\\
\hline\end{tabular}}\hfil
\hbox{\begin{tabular}{|r|}\hline
68\\70\\74\\106\\110\\
\hline\end{tabular}}\hfil
\hbox{\begin{tabular}{|r|}\hline
130\\142\\146\\154\\170\\
\hline\end{tabular}}\hfil
\hbox{\begin{tabular}{|r|}\hline
172\\212\\230\\234\\242\\
\hline\end{tabular}}\hfil
\hbox{\begin{tabular}{|r|}\hline
248\\266\\270\\286\\294\\
\hline\end{tabular}}\hfil
\hbox{\begin{tabular}{|r|}\hline
316\\318\\320\\322\\352\\
\hline\end{tabular}}\hfil
\hbox{\begin{tabular}{|r|}\hline
354\\356\\364\\380\\450\\
\hline\end{tabular}}\hfil
\hbox{\begin{tabular}{|r|}\hline
476\\510\\512\\648\\840\\
\hline\end{tabular}}\\[3mm]
Table II: Euler numbers not yet found in \cite{cls}

Apart from one exception, all of our 3-generation models come from potentials
with 5 points. They are given in tables III and IV.
The remaining 3 generation model comes from a configuration with 7 variables
requiring links:
$\{n_i\}=\{3, 4, 6, 6, 7, 7, 9\}$ and  $d=21$ with $\ngb=19$, $n_g=16$ and
$\c=6$.

{}~\\
\begin{tabular}{||rrrrr|r||rr|r||}
\hline\hline
$n_1$&$n_2$&$n_3$&$n_4$&$n_5$&$ d$ &$n_g$&$\bar n_g$&$\c$\\
\hline
4&4&5&5&7&25&17&20&-6\\
5&8&9&11&12&45&16&13&6\\
4&7&9&10&15&45&13&16&-6\\
3&5&8&14&15&45&20&23&-6\\
2&6&9&17&17&51&34&31&6\\
4&4&11&17&19&55&21&24&-6\\
3&4&14&21&21&63&32&35&-6\\
10&12&13&15&25&75&20&17&6\\
5&8&12&15&35&75&30&27&6\\
3&5&8&24&35&75&40&43&-6\\
2&9&19&24&27&81&32&29&6\\
3&7&18&26&27&81&29&32&-6\\
3&8&21&30&31&93&29&32&-6\\
1&13&23&28&32&97&47&50&-6\\
1&16&23&29&34&103&50&47&6\\
1&21&30&38&45&135&50&47&6\\
1&18&32&39&45&135&47&50&-6\\
5&6&14&45&65&135&45&42&6\\
2&8&29&49&59&147&48&51&-6\\
4&5&26&65&95&195&70&67&6\\
\hline\hline \end{tabular}\hfill
\hbox{\begin{tabular}{||rrrrr|r||rr|r||}
\hline\hline
$n_1$&$n_2$&$n_3$&$n_4$&$n_5$& d &$n_g$&$\bar n_g$&$\c$\\
\hline
3&4&6&13&13&39&29&26&6\\
4&4&7&13&15&43&20&23&-6\\
4&5&7&8&19&43&20&23&-6\\
3&5&8&9&20&45&23&26&-6\\
4&5&10&11&19&49&23&20&6\\
5&6&9&14&17&51&18&15&6\\
5&6&9&10&21&51&23&20&6\\
3&4&12&17&19&55&26&29&-6\\
3&12&15&16&17&63&29&32&-6\\
5&12&13&15&20&65&23&26&-6\\
3&9&17&22&24&75&35&38&-6\\
3&8&13&15&36&75&29&32&-6\\
2&6&15&23&31&77&37&40&-6\\
3&6&8&23&37&77&37&40&-6\\
2&9&12&23&37&83&40&37&6\\
4&6&15&35&45&105&37&34&6\\
4&7&13&41&58&123&40&37&6\\
3&10&18&59&87&177&57&60&-6\\
3&8&30&79&117&237&77&74&6\\
&&&&&&&&\\
\hline\hline
\end{tabular} }\\[3mm]
\hfil Table III: 3 generation models without links \hfill
Table IV: 5-pt. 3 gen. models requiring links

All candidates for 3 generation models belong to
polynomials in 5 or 7 variables, whereas the 2 generation models come
from configurations with 4 -- 6 non-trivial variables.
Surprisingly, there is exactly one 1 generation model, namely the model
we analysed in example 1: $\{n_i\}=\{5, 6, 7, 8, 9, 11, 12\}$ and $d=29$
with  $\ngb=13$, $n_g=12$ and $\c=2$.

Table V shows the models with the lowest total numbers of particles.
It is remarkable that they all belong to configurations with more than 5
variables. The 5 variable models with the lowest value of $n_g+\ngb$ are
the mirror pair of three generation models from table II with spectra
(16, 13; 6) and (13, 16; -6).

\noindent
\begin{tabular}{||rrrrrrrr|r||rr|rc||} \hline\hline
$n_1$&$n_2$&$n_3$&$n_4$&$n_5$&$n_6$&$n_7$&$n_8$&$d$&$n_g$&$\bar n_g$&
        $\c$&$n_g+\ngb$\\
\hline
5& 11& 12& 14& 18& 21&&& 54& 10& 10& 0& 20\\
7& 11& 12& 16& 20& 24&&& 60& 14& 8& 12& 22\\
6& 7& 9& 13 & 16& 21&&& 48& 11& 11& 0& 22\\
7& 18& 22& 24& 30& 34&&& 90& 11& 11& 0 &22\\
7& 9& 11& 12& 15& 17& 19&& 45& 11& 11& 0 &22\\
4& 4& 5& 5& 6& 7& 7&& 19& 8& 17& -18& 25\\
5& 6& 7& 8& 9& 11& 12&& 29& 13& 12& 2& 25\\
4& 6& 7& 9& 10& 15&&& 34& 7& 19& -24& 26\\
3& 7& 8& 10& 14& 15&&& 38& 8& 18& -20& 26\\
6& 10& 11& 13& 15& 20&&& 50& 17& 9& 16& 26\\
7& 7& 8& 8& 12& 12&&& 36& 16& 10& 12& 26\\
8& 9& 12& 17& 20& 24&&& 60& 19& 7& 24& 26\\
7& 10& 12& 16& 20& 25&&& 60& 7& 19& -24& 26\\
5& 5& 6& 6& 8& 9& 9&& 24& 13& 13& 0& 26\\
6& 9& 13& 14& 16& 17& 21&& 48& 13& 13& 0& 26\\
8& 8& 11& 11& 12& 12& 14& 14& 36& 16& 10& 12& 26\\
\hline\hline                 \end{tabular} \\[3mm]
Table V: The  models with lowest total numbers of particles

\noindent
In table VI we list the spectra which appear 20 or more times in our
list of (irreducible) non-degenerate models. All Euler numbers
among these spectra are multiples of 24. We have also given the sum of all
Hodge numbers $\bar \chi=P(1,1)=2(n_g+\ngb+2)$. It turns out that this number
also tends to being very ``non-prime''. In addition there are 88
configurations with $n_g=\bar n_g=21$ and $p_{01}=1$.

\noindent
\begin{tabular}{||rr|rr|r||}
\hline\hline
$n_g$&$\bar n_g$&$\c$&$\bar\chi$&\#\\
\hline
19&67&-96&176&29\\
8&44&-72&108&20\\
13&49&-72&128&28\\
11&35&-48&96&24\\
15&39&-48&112&20\\
23&47&-48&144&23\\
20&32&-24&108&22\\
19&19&0&80&38\\
23&23&0&96&38\\
29&29&0&120&27\\
31&31&0&128&39\\
39&39&0&160&39\\
43&43&0&176&26\\
\hline\hline \end{tabular}\hfill
\hbox{\begin{tabular}{||rr|rr|r||}
\hline\hline
$n_g$&$\bar n_g$&$\c$&$\bar\chi$&\#\\
\hline
53&53&0&216&38\\
55&55&0&224&32\\
29&17&24&96&26\\
32&20&24&108&29\\
35&11&48&96&31\\
39&15&48&112&20\\
41&17&48&120&27\\
43&19&48&128&26\\
44&8&72&108&27\\
46&10&72&116&34\\
49&13&72&128&38\\
56&20&72&156&25\\
77&41&72&240&22\\
\hline\hline \end{tabular}}\hfill
\hbox{\begin{tabular}{||rr|rr|r||}
\hline\hline
$n_g$&$\bar n_g$&$\c$&$\bar\chi$&\#\\
\hline
55&7&96&128&24\\
63&15&96&160&23\\
65&17&96&168&31\\
67&19&96&176&38\\
71&23&96&192&25\\
68&8&120&156&26\\
84&24&120&220&20\\
91&19&144&224&22\\
95&23&144&240&21\\
101&5&192&216&20\\
129&9&240&280&20\\
143&23&240&336&20\\
151&7&288&320&21\\
\hline\hline
\end{tabular} }\\[3mm]
Table VI: Spectra appearing 20 or more times \hfill

As a by-product of our classification of $\ch=3$ models we have also obtained
all $\ch=2$ configurations. It turns out that among these 124 models (which
have 3 to 6 variables) there is exactly one model, $\{n_i\}=\{3,3,4,4,4\}$,
$d=12$, with $b_{01}=2$ and $b_{11}=4$, whereas all other models have
$b_{01}=0$
and $b_{11}=20$. The first model factorizes and thus corresponds to a torus,
whereas all others have the same Hodge-diamond as the K3-surface~\cite{gh}.
This result is in fact known for the 17 models in our list which correspond to
minimal models \cite{g} and can be understood from the fact that,
topologically,
the only 2-dimensional complex manifolds of $SU(2)$ holonomy are the torus and
the K3 surface.
It also implies that the only case in which formula (\ref{xbar}) need not
apply is when $\chi=0$ and $P(1,1)=96$: For $D=3$ complete factorization, i.e.
a
3-dimensional torus, cannot occur without further orbifolding, as for $D=1$
the only possible values of $d$ are 3, 4 and 6.

\section{Discussion}

The most striking result of our calculations is definitely the strong
violation of mirror symmetry by non-invertible models. Whereas mirror
symmetry turned out to be nearly exact for invertible models, which were
already known to have mirror partners at least as orbifolds due to the
Berglund-H"ubsch-(BH-)construction \cite{bh}, half of the
non-degenerate models requiring links
turned out not to have mirror partners. This fact makes it seem
doubtful that all non-degenerate models have mirrors that can be described
as Landau-Ginzburg orbifolds.

%It is also worth noting that
Exactly half of our 3 generation models,
namely 20 out of 40, require links, whereas among all non-degenerate models
only 1 out of 6 does so. Note that in \cite{cls} only about 1000 configurations
of the 7555 Calabi-Yau's in weighted $\IP_4$ were missing, whereas these
authors only found 25 of the 39 models with $|\chi|=6$ in this class.
%This shows that 3 generations models tend to be represented by configurations
%requiring complicated links.
With the help of the BH-construction we can easily
construct at least 9 further 3 generation models as orbifolds of the invertible
models of table III (3 of them are already given in \cite{kss}).

Our work might be useful for the construction of even
more 3 generation models due to the following considerations:
As already observed by Candelas et al. \cite{cls}, the Euler number is a
multiple of 12 (or even 24) in most cases. On the other hand, all known 3
generation
orbifolds come from potentials where the Euler number is an odd multiple
of 6. These configurations, of which we have 496 in our list, may be
the most promising for further investigation. A mirror pair of such
orbifolds is given in \cite{kss}.
The fact that all 3 generation models and candidates for 3 generation models
come from 5- or 7-point polynomials is a consequence of the fact that only
polynomials with an odd number of variables can have odd $d$
%(due to $\sum n_i=d(n-3)/2$)
and only models with odd $d$ allow Euler numbers which are not divisible by 4.

\old
Another point that should be stressed is the fact that with the present
result we are very close to fulfilling  a task that might at first sight
seem much bigger: the construction of all abelian orbifolds of Landau-Ginzburg
models. The way of achieving this is straightforward: Take any configuration
in our list, determine all skeleton graphs and all links that make them
non-degenerate, find all phase symmetries of the resulting polynomials and
orbifold with respect to them. One might object that this would not cover
abelian subgroups of possible permutation groups. By diagonalizing the
representations of such groups, however (at this point it is crucial that we
consider only abelian groups), we can always make them act as pure phase
symmetries.
\endold
Although we have restricted attention to the canonical $\ZZ_d$ orbifold, our
results yield an implicit classification of all abelian orbifolds, because any
abelian group action can be diagonalized. Thus we can assume that such a
group acts as a phase symmetry and a calculation of all possible systems of
links generates all points of maximal abelian symmetry.
This procedure may take us to different points
in moduli space, as the following example illustrates: Consider
$W=X_1^3+X_2^3$
and the symmetry $X_1 \leftrightarrow X_2$, which acts diagonally
on $\tilde X_1=X_1+X_2$ and $\tilde X_2=X_1-X_2$.
In these new variables $W=\tilde X_1^3/4+3\tilde X_1\tilde X_2^2/4$.

Several questions are raised by our work:
Are there interpretations of all concepts of singularity theory,
specifically of our link criterion, in terms of N=2 SCFT?
What is the role of the almost non-degenerate configurations: Could we
conclude from the applicability of formulas \ref{sum} and \ref{vaf}
that almost non-degenerate configurations correspond to genuine SCFTs?
%(Unfortunately,
(Including their spectra would not improve MS: Instead of
providing partners we would rather generate new singles.)
Most pressing, however, is the question where we should look
for the mirror partners of the non-invertible models.

{\it Acknowledgements:} We are indebted to Albrecht Klemm for communicating
results of his related calculations~\cite{kle} prior to publication.
M.K. would also like to thank Per Berglund for interesting discussions on
his work~\cite{bh}.
\ve

\ifnum\npre=2 \end{document}\fi  \cut{\let\ve=\vfil}
\newpage \noindent {\Large\bf Figures}  \pagestyle{empty}  \bsk\bsk
\ifnum\npre=7 \def\hocut#1{\hopl}  \def\hocucut#1{\hocut}
      \else \def\hocut#1#2{\hopl{#1#2}}\def\hocucut#1#2#3{\hopl{#1#2#3}} \fi

\def\figa{\he=3500 \vdiv=4  \ptsize=2 \def\vq{$d$} \def\hq{$\bar n_g$}
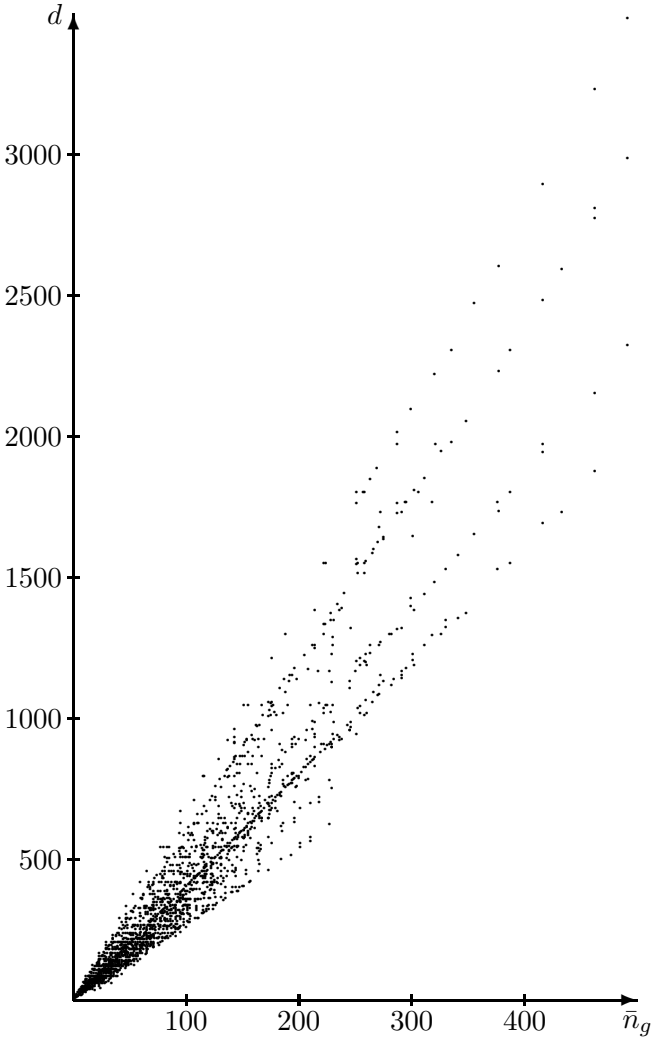
\def\capt{Fig. 1: A plot of $d$ vs. the number of $\bar{27}$
	representations for 4 variables.}
\def\vertlab{\vlab{500} \vlab{1000} \vlab{1500}\vlab{2000}\vlab{2500}
	      \vlab{3000}} %\hopl{{\input d-ngb-4 }}
}
\def\figb{\he=2000 \vdiv=4  \ptsize=2 \def\vq{$d$} \def\hq{$\bar n_g$}
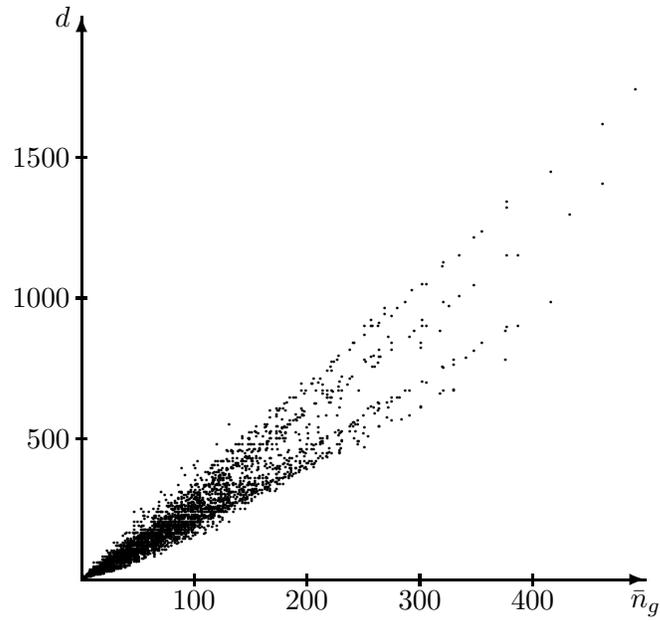
\def\capt{Fig. 2: A plot of $d$ vs. the number of $\bar{27}$
	representations for 5 variables.}
\def\vertlab{\vlab{500} \vlab{1000} \vlab{1500}} %\hopl{{\input d-ngb-5 }}
}
\def\figc{\he=2000 \vdiv=4  \ptsize=2 \def\vq{$d$} \def\hq{$\bar n_g$}
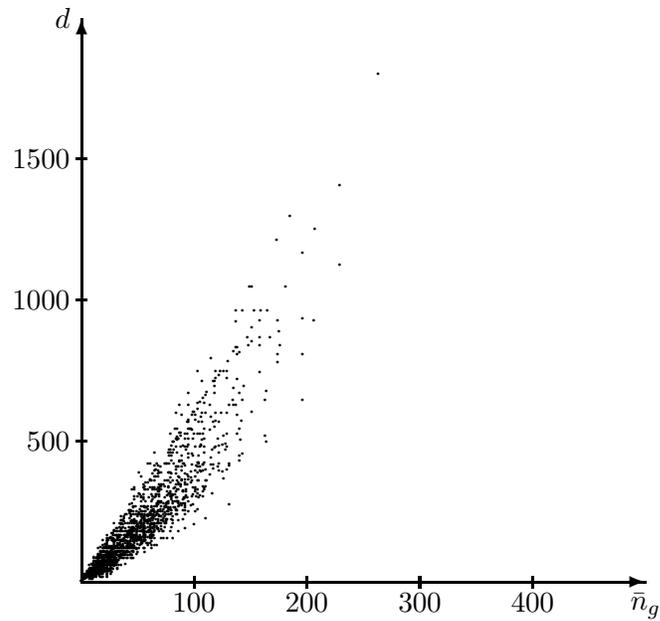
\def\capt{Fig. 3: A plot of $d$ vs. the number of $\bar{27}$
	representations for 6 variables.}
\def\vertlab{\vlab{500} \vlab{1000} \vlab{1500}} %\hopl{{\input d-ngb-6 }}
}
\def\figd{\he=2000 \vdiv=4  \ptsize=2 \def\vq{$d$} \def\hq{$\bar n_g$}
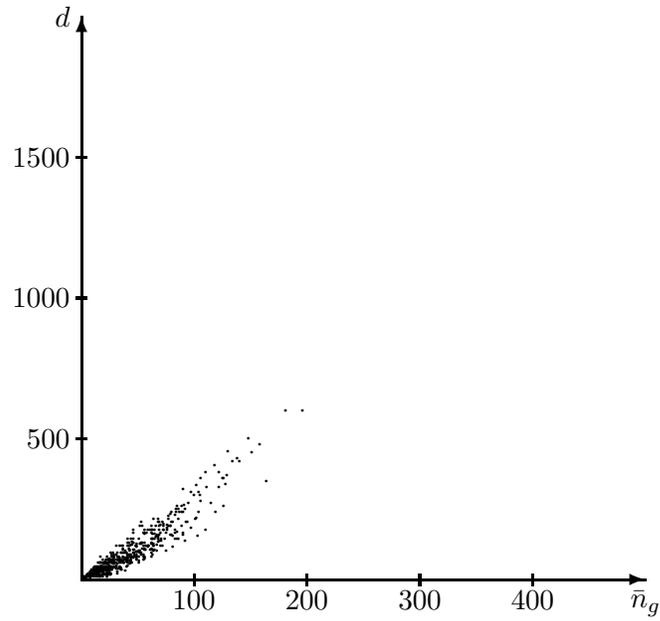
\def\capt{Fig. 4: A plot of $d$ vs. the number of $\bar{27}$
      representations for 7 variables.}
\def\vertlab{\vlab{500} \vlab{1000} \vlab{1500}} %\hopl{{\input d-ngb-7 }}
}
\def\fige{\def\.##1,##2){\put(##2,##1){\circle*{2}}}
\he=500 \vdiv=1  \def\vq{$p$} \def\hq{$\bar n_g$}
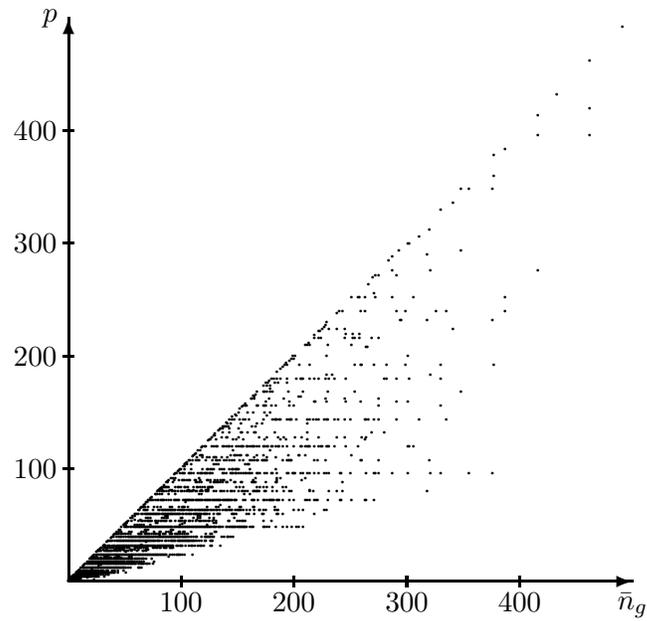
\def\capt{Fig. 5: A plot of $p=\phi(d)/2-1$ vs. the number of $\bar{27}$
	representations for all non-degenerate configurations.}
\def\vertlab{\vlab{100}\vlab{200}\vlab{300}\vlab{400}} %\hopl{{\input p-ngb }}
}

\figa  \hocucut{
\.8,1)\.10,1)\.12,2)\.12,3)\.14,2)\.16,3)\.16,4)
\.18,2)\.18,3)\.18,4)\.20,3)\.20,4)\.20,6)\.20,7)
\.22,4)\.24,3)\.24,6)\.24,8)\.26,5)\.28,5)\.28,6)
\.28,8)\.28,9)\.28,11)\.30,4)\.30,5)\.30,6)\.30,7)
\.30,8)\.30,9)\.32,7)\.32,8)\.32,9)\.32,10)\.34,7)
\.36,5)\.36,7)\.36,8)\.36,9)\.36,10)\.36,11)\.36,12)
\.36,19)\.38,8)\.40,7)\.40,9)\.40,10)\.40,11)\.40,12)
\.40,13)\.40,14)\.40,15)\.42,6)\.42,7)\.42,8)\.42,9)
\.42,11)\.42,12)\.42,17)\.44,10)\.44,11)\.44,14)\.44,15)
\.44,24)\.46,10)\.48,7)\.48,8)\.48,9)\.48,12)\.48,13)
\.48,14)\.48,15)\.48,16)\.50,9)\.50,11)\.50,13)\.50,17)
\.52,12)\.52,14)\.52,15)\.52,17)\.52,18)\.52,23)\.54,8)
\.54,10)\.54,11)\.54,12)\.54,14)\.54,15)\.54,19)\.56,11)
\.56,13)\.56,14)\.56,16)\.56,17)\.56,19)\.56,23)\.58,13)
\.60,10)\.60,11)\.60,12)\.60,13)\.60,15)\.60,16)\.60,17)
\.60,18)\.60,19)\.60,20)\.60,22)\.60,25)\.60,26)\.60,28)
\.60,34)\.62,14)\.64,15)\.64,16)\.64,18)\.64,19)\.64,21)
\.64,22)\.66,9)\.66,10)\.66,11)\.66,14)\.66,15)\.66,16)
\.66,19)\.66,24)\.66,29)\.68,16)\.68,17)\.68,18)\.68,19)
\.68,23)\.68,24)\.68,31)\.70,11)\.70,13)\.70,14)\.70,15)
\.70,16)\.70,17)\.70,18)\.70,19)\.70,20)\.70,23)\.70,25)
\.70,26)\.70,27)\.72,11)\.72,14)\.72,15)\.72,16)\.72,17)
\.72,18)\.72,19)\.72,20)\.72,21)\.72,22)\.72,23)\.72,26)
\.74,17)\.76,17)\.76,20)\.76,26)\.76,27)\.76,32)\.78,11)
\.78,12)\.78,17)\.78,18)\.78,21)\.78,23)\.78,24)\.78,27)
\.80,15)\.80,19)\.80,21)\.80,22)\.80,23)\.80,24)\.80,25)
\.80,26)\.80,28)\.80,29)\.80,33)\.82,19)\.84,11)\.84,13)
\.84,14)\.84,15)\.84,16)\.84,17)\.84,20)\.84,23)\.84,24)
\.84,26)\.84,28)\.84,29)\.84,30)\.84,32)\.84,34)\.84,38)
\.86,20)\.88,22)\.88,23)\.88,24)\.88,26)\.88,27)\.88,29)
\.88,31)\.88,34)\.90,14)\.90,15)\.90,16)\.90,17)\.90,18)
\.90,19)\.90,20)\.90,21)\.90,22)\.90,23)\.90,25)\.90,26)
\.90,27)\.90,29)\.90,33)\.90,34)\.90,35)\.90,36)\.90,39)
\.92,22)\.92,23)\.92,25)\.92,33)\.92,36)\.92,43)\.94,22)
\.96,16)\.96,18)\.96,23)\.96,24)\.96,25)\.96,27)\.96,29)
\.96,31)\.96,32)\.96,33)\.96,34)\.98,20)\.98,23)\.98,29)
\.100,24)\.100,27)\.100,28)\.100,29)\.100,31)\.100,33)\.100,34)
\.100,36)\.100,39)\.102,15)\.102,16)\.102,23)\.102,24)\.102,31)
\.102,32)\.102,39)\.104,23)\.104,25)\.104,26)\.104,29)\.104,31)
\.104,37)\.106,25)\.108,18)\.108,19)\.108,24)\.108,25)\.108,26)
\.108,27)\.108,30)\.108,31)\.108,32)\.108,33)\.108,38)\.108,41)
\.110,21)\.110,24)\.110,26)\.110,28)\.110,29)\.110,39)\.110,41)
\.112,23)\.112,27)\.112,29)\.112,30)\.112,32)\.112,33)\.112,34)
\.112,35)\.112,36)\.112,37)\.112,38)\.114,17)\.114,18)\.114,19)
\.114,26)\.114,27)\.114,29)\.114,32)\.114,35)\.116,28)\.116,31)
\.116,34)\.116,42)\.116,46)\.118,28)\.120,19)\.120,20)\.120,22)
\.120,23)\.120,24)\.120,25)\.120,27)\.120,29)\.120,30)\.120,31)
\.120,32)\.120,33)\.120,34)\.120,35)\.120,36)\.120,37)\.120,38)
\.120,39)\.120,40)\.120,41)\.120,43)\.120,44)\.120,47)\.122,29)
\.124,32)\.124,34)\.124,39)\.124,47)\.124,49)\.124,54)\.126,17)
\.126,21)\.126,22)\.126,23)\.126,26)\.126,27)\.126,28)\.126,29)
\.126,30)\.126,31)\.126,32)\.126,33)\.126,34)\.126,35)\.126,38)
\.126,41)\.126,42)\.128,33)\.128,43)\.128,45)\.128,46)\.130,23)
\.130,27)\.130,29)\.130,31)\.130,35)\.130,37)\.130,41)\.130,47)
\.130,49)\.130,51)\.132,21)\.132,22)\.132,23)\.132,33)\.132,34)
\.132,35)\.132,39)\.132,40)\.132,46)\.132,47)\.132,48)\.134,32)
\.136,33)\.136,37)\.136,39)\.136,41)\.136,49)\.138,22)\.138,32)
\.138,33)\.138,34)\.138,43)\.140,23)\.140,28)\.140,29)\.140,32)
\.140,33)\.140,34)\.140,35)\.140,36)\.140,37)\.140,38)\.140,39)
\.140,40)\.140,44)\.140,45)\.140,46)\.140,47)\.140,49)\.140,50)
\.140,51)\.140,54)\.140,57)\.142,34)\.144,23)\.144,31)\.144,32)
\.144,35)\.144,36)\.144,38)\.144,41)\.144,43)\.144,47)\.144,49)
\.144,50)\.144,53)\.146,35)\.148,38)\.148,54)\.148,56)\.148,59)
\.150,23)\.150,24)\.150,27)\.150,28)\.150,29)\.150,31)\.150,33)
\.150,39)\.152,35)\.152,41)\.152,48)\.152,50)\.154,29)\.154,32)
\.154,37)\.154,39)\.154,42)\.154,47)\.154,49)\.156,23)\.156,25)
\.156,26)\.156,28)\.156,38)\.156,41)\.156,44)\.156,53)\.156,55)
\.156,56)\.156,57)\.158,38)\.160,33)\.160,39)\.160,41)\.160,45)
\.160,46)\.160,48)\.160,50)\.160,51)\.160,53)\.160,55)\.160,57)
\.162,26)\.162,34)\.162,38)\.162,39)\.162,43)\.162,51)\.164,40)
\.164,43)\.164,64)\.168,23)\.168,28)\.168,30)\.168,33)\.168,34)
\.168,35)\.168,43)\.168,46)\.168,47)\.168,51)\.168,53)\.168,56)
\.168,57)\.168,59)\.168,61)\.170,33)\.170,37)\.170,39)\.170,41)
\.170,47)\.170,49)\.170,65)\.172,42)\.172,43)\.172,45)\.172,48)
\.174,27)\.174,48)\.176,43)\.176,45)\.176,46)\.176,47)\.176,54)
\.176,59)\.176,60)\.176,65)\.180,29)\.180,31)\.180,32)\.180,35)
\.180,37)\.180,40)\.180,41)\.180,42)\.180,44)\.180,47)\.180,50)
\.180,51)\.180,52)\.180,54)\.180,55)\.180,56)\.180,57)\.180,59)
\.180,62)\.180,65)\.180,66)\.182,35)\.182,38)\.182,44)\.182,50)
\.182,56)\.182,59)\.184,43)\.184,45)\.184,56)\.186,29)\.186,30)
\.186,39)\.186,46)\.186,49)\.186,54)\.186,60)\.186,61)\.186,65)
\.188,46)\.188,71)\.190,35)\.190,37)\.190,39)\.190,44)\.190,46)
\.190,48)\.190,53)\.190,71)\.192,31)\.192,39)\.192,53)\.192,56)
\.192,59)\.192,61)\.192,64)\.192,66)\.192,68)\.192,71)\.194,47)
\.196,42)\.196,47)\.196,55)\.196,62)\.196,69)\.196,76)\.198,29)
\.198,32)\.198,34)\.198,42)\.198,43)\.198,45)\.198,46)\.198,52)
\.198,59)\.198,60)\.198,61)\.198,63)\.200,43)\.200,49)\.200,51)
\.200,55)\.200,56)\.200,57)\.200,61)\.200,65)\.200,67)\.200,69)
\.200,71)\.202,49)\.204,35)\.204,50)\.204,51)\.204,61)\.204,62)
\.204,71)\.204,75)\.206,50)\.208,59)\.208,64)\.208,66)\.208,69)
\.208,71)\.208,77)\.210,29)\.210,34)\.210,36)\.210,37)\.210,39)
\.210,40)\.210,41)\.210,42)\.210,43)\.210,44)\.210,45)\.210,46)
\.210,47)\.210,48)\.210,50)\.210,53)\.210,55)\.210,57)\.210,59)
\.210,62)\.210,65)\.210,69)\.212,79)\.214,52)\.216,35)\.216,44)
\.216,48)\.216,49)\.216,51)\.216,52)\.216,54)\.216,57)\.216,59)
\.216,61)\.216,62)\.216,63)\.216,65)\.216,68)\.218,53)\.220,45)
\.220,47)\.220,56)\.220,58)\.220,59)\.220,65)\.220,67)\.220,68)
\.220,69)\.220,74)\.220,75)\.220,79)\.220,81)\.220,82)\.220,84)
\.222,35)\.222,56)\.222,72)\.222,75)\.224,56)\.224,61)\.224,66)
\.224,68)\.224,71)\.224,76)\.224,79)\.224,82)\.228,38)\.228,62)
\.228,65)\.228,68)\.228,71)\.228,83)\.228,86)\.228,91)\.230,43)
\.230,67)\.230,69)\.232,57)\.232,69)\.232,71)\.234,35)\.234,38)
\.234,41)\.234,51)\.234,53)\.234,59)\.234,63)\.234,65)\.234,71)
\.236,88)\.238,50)\.238,56)\.238,63)\.238,71)\.238,77)\.240,39)
\.240,40)\.240,44)\.240,46)\.240,48)\.240,50)\.240,51)\.240,53)
\.240,55)\.240,59)\.240,63)\.240,65)\.240,67)\.240,68)\.240,73)
\.240,75)\.240,79)\.240,81)\.240,84)\.240,87)\.240,88)\.242,64)
\.244,60)\.244,90)\.244,95)\.246,64)\.250,65)\.250,73)\.252,43)
\.252,55)\.252,59)\.252,65)\.252,68)\.252,70)\.252,72)\.252,76)
\.252,80)\.252,83)\.252,91)\.252,93)\.254,62)\.256,69)\.256,87)
\.256,91)\.256,93)\.256,94)\.258,41)\.258,45)\.258,48)\.258,63)
\.260,52)\.260,60)\.260,65)\.260,70)\.260,73)\.260,77)\.260,83)
\.260,89)\.260,90)\.260,94)\.260,100)\.264,46)\.264,54)\.264,57)
\.264,76)\.264,81)\.264,83)\.264,85)\.264,89)\.264,95)\.266,68)
\.266,71)\.266,80)\.266,83)\.268,66)\.268,67)\.268,68)\.268,98)
\.270,44)\.270,48)\.270,53)\.270,54)\.270,55)\.270,57)\.270,60)
\.270,62)\.270,63)\.270,67)\.270,69)\.270,71)\.270,72)\.270,77)
\.270,81)\.270,86)\.272,75)\.272,87)\.272,91)\.272,92)\.276,43)
\.276,47)\.276,98)\.276,102)\.276,103)\.278,68)\.280,59)\.280,67)
\.280,72)\.280,73)\.280,77)\.280,78)\.280,82)\.280,84)\.280,86)
\.280,88)\.280,92)\.280,101)\.282,68)\.282,69)\.286,70)\.286,81)
\.288,47)\.288,50)\.288,74)\.288,77)\.288,81)\.288,86)\.288,91)
\.288,95)\.288,98)\.288,103)\.290,69)\.290,71)\.290,83)\.292,77)
\.292,108)\.292,110)\.292,111)\.294,41)\.294,47)\.294,48)\.294,51)
\.294,55)\.294,62)\.294,77)\.294,81)\.296,91)\.296,95)\.298,73)
\.300,59)\.300,60)\.300,71)\.300,74)\.300,79)\.300,86)\.300,87)
\.300,92)\.300,94)\.300,104)\.302,74)\.304,75)\.304,84)\.304,96)
\.304,98)\.304,102)\.304,103)\.306,47)\.306,64)\.306,66)\.306,69)
\.306,73)\.306,83)\.308,79)\.308,88)\.308,89)\.308,97)\.308,101)
\.308,114)\.308,115)\.310,59)\.310,78)\.310,80)\.310,86)\.310,91)
\.310,95)\.312,53)\.312,64)\.312,66)\.312,81)\.312,83)\.312,91)
\.312,95)\.312,116)\.314,77)\.316,78)\.316,81)\.316,118)\.318,51)
\.318,52)\.318,79)\.320,69)\.320,85)\.320,86)\.320,87)\.320,91)
\.320,93)\.320,94)\.320,97)\.320,109)\.320,115)\.322,68)\.322,71)
\.322,79)\.322,101)\.324,71)\.324,73)\.324,83)\.324,87)\.324,98)
\.324,118)\.324,120)\.324,124)\.326,80)\.328,81)\.328,89)\.328,101)
\.328,103)\.330,54)\.330,58)\.330,59)\.330,63)\.330,64)\.330,67)
\.330,68)\.330,69)\.330,71)\.330,76)\.330,91)\.330,109)\.332,83)
\.334,82)\.336,47)\.336,55)\.336,63)\.336,64)\.336,65)\.336,71)
\.336,86)\.336,90)\.336,97)\.336,101)\.336,109)\.336,115)\.336,119) }{
\.340,70)\.340,81)\.340,88)\.340,90)\.340,99)\.340,100)\.340,102)
\.340,120)\.340,124)\.340,126)\.342,71)\.342,74)\.342,78)\.342,86)
\.342,91)\.342,98)\.342,99)\.342,101)\.342,103)\.344,127)\.346,85)
\.350,71)\.350,72)\.350,74)\.350,77)\.350,79)\.350,84)\.350,86)
\.350,92)\.350,95)\.352,79)\.352,92)\.352,94)\.352,95)\.352,100)
\.352,108)\.354,87)\.356,133)\.360,61)\.360,71)\.360,75)\.360,77)
\.360,78)\.360,79)\.360,81)\.360,86)\.360,94)\.360,97)\.360,98)
\.360,99)\.360,100)\.360,101)\.360,102)\.360,103)\.360,104)\.360,108)
\.360,110)\.360,111)\.360,113)\.362,89)\.364,80)\.364,83)\.364,85)
\.364,95)\.364,97)\.364,116)\.364,121)\.364,135)\.366,60)\.366,61)
\.366,95)\.366,120)\.368,93)\.368,126)\.368,131)\.370,73)\.370,75)
\.370,97)\.372,62)\.372,95)\.372,98)\.374,84)\.374,97)\.376,93)
\.376,95)\.376,116)\.378,53)\.378,62)\.378,68)\.378,70)\.378,72)
\.378,82)\.378,83)\.378,85)\.378,91)\.378,100)\.378,116)\.380,114)
\.380,115)\.380,125)\.380,135)\.384,103)\.384,117)\.384,140)\.384,141)
\.386,95)\.388,96)\.388,99)\.388,145)\.388,146)\.390,71)\.390,75)
\.390,77)\.390,81)\.390,83)\.390,100)\.390,101)\.392,83)\.392,106)
\.392,121)\.392,125)\.394,97)\.396,68)\.396,89)\.396,92)\.396,104)
\.396,105)\.396,107)\.396,109)\.396,112)\.396,122)\.398,98)\.400,119)
\.400,135)\.400,141)\.400,144)\.400,147)\.402,66)\.402,68)\.404,150)
\.404,151)\.406,86)\.406,89)\.406,100)\.408,67)\.408,70)\.408,86)
\.408,99)\.408,105)\.408,121)\.410,81)\.410,89)\.410,123)\.412,102)
\.414,71)\.414,92)\.414,103)\.414,123)\.414,131)\.416,95)\.416,103)
\.416,109)\.416,115)\.416,128)\.418,94)\.420,59)\.420,67)\.420,71)
\.420,76)\.420,78)\.420,82)\.420,84)\.420,87)\.420,90)\.420,91)
\.420,92)\.420,95)\.420,107)\.420,110)\.420,116)\.420,118)\.420,119)
\.420,122)\.420,123)\.420,124)\.420,125)\.420,127)\.420,156)\.422,104)
\.424,133)\.424,157)\.428,105)\.430,125)\.432,79)\.432,80)\.432,107)
\.432,117)\.432,122)\.432,128)\.432,133)\.434,98)\.434,107)\.436,108)
\.438,72)\.438,115)\.440,79)\.440,107)\.440,109)\.440,113)\.440,114)
\.440,117)\.440,124)\.440,131)\.440,136)\.440,144)\.442,101)\.442,103)
\.442,109)\.442,115)\.442,117)\.444,74)\.446,110)\.448,95)\.448,111)
\.448,117)\.448,118)\.448,123)\.448,153)\.450,77)\.450,81)\.450,89)
\.450,98)\.450,99)\.450,101)\.450,104)\.450,111)\.450,117)\.450,122)
\.450,123)\.454,112)\.456,95)\.456,96)\.456,143)\.460,93)\.460,114)
\.460,116)\.460,131)\.460,137)\.460,141)\.460,164)\.462,65)\.462,79)
\.462,81)\.462,85)\.462,87)\.462,89)\.462,101)\.462,107)\.462,114)
\.462,118)\.464,119)\.464,144)\.464,153)\.464,173)\.468,106)\.468,122)
\.468,128)\.468,134)\.468,144)\.470,95)\.470,116)\.472,117)\.474,81)
\.474,116)\.476,102)\.476,122)\.478,118)\.480,86)\.480,88)\.480,92)
\.480,94)\.480,105)\.480,109)\.480,117)\.480,131)\.480,141)\.484,123)
\.484,131)\.484,165)\.486,107)\.486,111)\.486,118)\.486,124)\.488,127)
\.490,97)\.490,103)\.490,124)\.492,83)\.492,122)\.494,116)\.494,122)
\.496,138)\.496,149)\.496,164)\.500,184)\.504,83)\.504,85)\.504,122)
\.504,123)\.504,130)\.504,131)\.506,114)\.506,125)\.506,131)\.506,136)
\.508,126)\.510,81)\.510,90)\.510,95)\.510,97)\.510,99)\.510,100)
\.510,102)\.510,103)\.510,113)\.510,115)\.510,120)\.510,131)\.510,145)
\.516,193)\.518,116)\.518,128)\.520,95)\.520,103)\.520,137)\.520,163)
\.522,89)\.522,114)\.522,115)\.522,119)\.526,130)\.528,94)\.528,100)
\.528,108)\.528,138)\.528,151)\.528,165)\.530,131)\.532,115)\.532,116)
\.532,132)\.532,138)\.532,143)\.534,133)\.536,166)\.540,94)\.540,95)
\.540,98)\.540,114)\.540,124)\.540,140)\.540,163)\.540,164)\.544,136)
\.544,141)\.544,151)\.544,201)\.546,77)\.546,80)\.546,83)\.546,85)
\.546,96)\.546,97)\.546,105)\.546,109)\.546,118)\.546,119)\.546,126)
\.546,129)\.546,131)\.546,145)\.550,109)\.550,124)\.550,134)\.550,136)
\.550,138)\.550,143)\.550,145)\.550,146)\.552,113)\.552,116)\.552,153)
\.552,166)\.556,139)\.558,126)\.558,137)\.558,164)\.560,117)\.560,176)
\.560,201)\.564,210)\.568,141)\.570,102)\.570,107)\.570,117)\.570,132)
\.572,132)\.572,143)\.574,145)\.576,119)\.576,151)\.576,157)\.578,143)
\.578,151)\.580,117)\.580,144)\.580,150)\.580,152)\.580,173)\.580,210)
\.582,146)\.588,98)\.588,104)\.588,106)\.588,128)\.588,152)\.588,155)
\.590,146)\.592,147)\.592,155)\.594,94)\.594,102)\.594,105)\.594,124)
\.594,130)\.594,132)\.594,135)\.594,144)\.596,148)\.600,119)\.600,123)
\.600,124)\.600,155)\.600,185)\.602,131)\.602,149)\.604,151)\.606,151)
\.608,143)\.608,157)\.610,121)\.610,127)\.612,137)\.612,144)\.612,161)
\.614,152)\.616,137)\.616,154)\.616,158)\.618,102)\.620,184)\.622,154)
\.624,107)\.624,128)\.624,227)\.628,158)\.630,95)\.630,107)\.630,110)
\.630,111)\.630,119)\.630,124)\.630,133)\.630,135)\.630,140)\.630,144)
\.630,155)\.630,186)\.632,196)\.634,157)\.636,159)\.636,160)\.638,167)
\.646,152)\.648,134)\.648,145)\.648,161)\.648,167)\.648,173)\.648,196)
\.650,129)\.650,131)\.650,165)\.650,167)\.650,173)\.652,162)\.652,165)
\.660,110)\.660,114)\.660,129)\.660,155)\.660,159)\.660,174)\.662,164)
\.664,165)\.666,147)\.666,155)\.668,166)\.672,95)\.672,118)\.672,149)
\.672,160)\.672,181)\.674,167)\.676,174)\.676,175)\.680,169)\.680,201)
\.682,159)\.682,164)\.682,179)\.684,143)\.684,152)\.686,149)\.686,176)
\.688,174)\.690,116)\.690,129)\.690,141)\.690,148)\.690,166)\.696,144)    }{
\.696,147)\.698,173)\.700,140)\.700,151)\.700,176)\.700,210)\.702,165)
\.702,172)\.702,173)\.704,196)\.704,218)\.708,179)\.712,177)\.714,107)
\.714,121)\.714,126)\.714,170)\.720,126)\.720,138)\.720,149)\.720,160)
\.720,174)\.720,218)\.722,179)\.726,123)\.726,165)\.726,176)\.726,184)
\.728,181)\.728,183)\.730,145)\.736,183)\.740,184)\.744,149)\.744,158)
\.750,161)\.752,229)\.756,131)\.756,173)\.756,197)\.758,188)\.760,143)
\.760,157)\.760,187)\.760,198)\.762,126)\.766,190)\.768,191)\.770,151)
\.770,186)\.772,194)\.774,174)\.774,193)\.780,174)\.780,206)\.780,227)
\.782,186)\.782,194)\.784,195)\.784,208)\.786,130)\.788,196)\.790,200)
\.792,137)\.796,200)\.798,115)\.798,116)\.798,134)\.798,138)\.798,143)
\.798,175)\.798,190)\.798,198)\.806,194)\.810,145)\.810,147)\.810,163)
\.810,176)\.810,194)\.810,195)\.814,202)\.816,136)\.822,137)\.826,176)
\.826,179)\.828,183)\.830,204)\.830,206)\.832,214)\.834,139)\.840,145)
\.840,148)\.840,155)\.840,158)\.840,176)\.840,185)\.840,208)\.844,210)
\.846,142)\.852,212)\.854,182)\.856,213)\.858,129)\.858,143)\.858,207)
\.864,143)\.864,214)\.870,144)\.870,148)\.870,150)\.870,152)\.870,159)
\.870,167)\.870,210)\.880,219)\.882,155)\.882,188)\.888,185)\.892,223)
\.900,164)\.900,194)\.900,214)\.900,229)\.902,229)\.904,225)\.906,151)
\.908,227)\.910,183)\.910,194)\.910,197)\.910,220)\.910,222)\.910,229)
\.912,143)\.916,228)\.918,143)\.918,152)\.924,137)\.924,154)\.924,161)
\.924,206)\.924,236)\.928,231)\.928,238)\.930,154)\.930,159)\.930,164)
\.930,169)\.930,194)\.930,200)\.936,143)\.936,196)\.936,214)\.936,236)
\.940,234)\.942,158)\.946,214)\.946,230)\.946,251)\.960,176)\.960,207)
\.960,245)\.962,245)\.966,143)\.966,212)\.968,246)\.972,163)\.976,243)
\.978,162)\.978,165)\.990,214)\.990,231)\.990,246)\.994,212)\.1008,173)
\.1012,258)\.1014,168)\.1014,175)\.1014,180)\.1020,186)\.1020,222)\.1020,260)
\.1022,254)\.1026,178)\.1026,228)\.1032,174)\.1036,222)\.1036,252)\.1044,176)
\.1044,254)\.1050,151)\.1050,155)\.1050,167)\.1050,177)\.1050,190)\.1050,210)
\.1050,224)\.1050,225)\.1050,230)\.1056,175)\.1056,218)\.1060,265)\.1062,173)
\.1062,176)\.1066,259)\.1084,270)\.1088,271)\.1092,188)\.1092,266)\.1110,183)
\.1110,245)\.1120,271)\.1120,282)\.1128,229)\.1134,191)\.1134,245)\.1134,275)
\.1140,187)\.1140,198)\.1142,284)\.1144,291)\.1156,272)\.1156,291)\.1158,192)
\.1158,194)\.1170,217)\.1170,227)\.1170,250)\.1170,291)\.1176,208)\.1180,295)
\.1182,196)\.1188,302)\.1190,254)\.1190,260)\.1200,259)\.1204,251)\.1204,258)
\.1208,301)\.1218,176)\.1218,254)\.1224,205)\.1230,229)\.1230,259)\.1230,301)
\.1232,263)\.1248,214)\.1260,212)\.1260,214)\.1260,230)\.1260,263)\.1260,271)
\.1260,311)\.1274,272)\.1290,230)\.1298,318)\.1300,326)\.1302,188)\.1302,222)
\.1302,280)\.1302,282)\.1316,287)\.1320,246)\.1320,291)\.1324,330)\.1338,222)
\.1338,223)\.1350,228)\.1350,231)\.1350,330)\.1356,341)\.1372,348)\.1374,228)
\.1386,214)\.1386,236)\.1386,302)\.1392,238)\.1400,299)\.1410,234)\.1428,299)
\.1442,311)\.1446,240)\.1484,320)\.1518,252)\.1518,258)\.1530,330)\.1534,376)
\.1548,251)\.1552,387)\.1554,222)\.1554,224)\.1554,252)\.1554,258)\.1560,259)
\.1566,251)\.1582,341)\.1590,265)\.1602,266)\.1626,270)\.1638,275)\.1644,275)
\.1650,301)\.1652,355)\.1680,271)\.1692,416)\.1728,287)\.1732,433)\.1734,272)
\.1734,291)\.1736,377)\.1764,251)\.1764,287)\.1770,294)\.1770,295)\.1770,318)
\.1770,376)\.1806,251)\.1806,257)\.1806,258)\.1806,306)\.1806,387)\.1812,302)
\.1848,263)\.1854,311)\.1876,462)\.1890,269)\.1946,416)\.1950,326)\.1974,287)
\.1974,321)\.1974,416)\.1980,335)\.2016,287)\.2058,348)\.2100,299)\.2156,462)
\.2226,320)\.2232,377)\.2310,335)\.2310,387)\.2324,491)\.2478,355)\.2484,416)
\.2598,433)\.2604,377)\.2778,462)\.2814,462)\.2898,416)\.2988,491)\.3234,462)
\.3486,491)} \ve

\figb  \hocucut{
\.5,1)\.6,1)\.7,2)\.8,2)\.9,2)\.9,3)\.9,4)
\.10,3)\.11,4)\.12,2)\.12,3)\.12,4)\.12,5)\.12,6)
\.13,5)\.14,5)\.15,3)\.15,4)\.15,5)\.15,6)\.15,7)
\.15,8)\.15,9)\.16,5)\.16,6)\.16,8)\.17,7)\.18,5)
\.18,6)\.18,7)\.18,9)\.18,11)\.19,8)\.20,5)\.20,6)
\.20,7)\.20,8)\.20,9)\.20,11)\.20,12)\.20,15)\.21,5)
\.21,6)\.21,7)\.21,8)\.21,9)\.21,10)\.21,11)\.21,12)
\.21,14)\.21,17)\.22,9)\.23,10)\.24,6)\.24,7)\.24,8)
\.24,9)\.24,10)\.24,11)\.24,12)\.24,13)\.24,14)\.24,15)
\.24,19)\.24,20)\.25,9)\.25,11)\.25,13)\.25,17)\.26,11)
\.27,8)\.27,9)\.27,10)\.27,11)\.27,12)\.27,13)\.27,14)
\.27,15)\.27,16)\.27,19)\.28,8)\.28,9)\.28,10)\.28,11)
\.28,12)\.28,13)\.28,14)\.28,17)\.28,20)\.28,23)\.29,13)
\.30,7)\.30,8)\.30,9)\.30,11)\.30,12)\.30,13)\.30,15)
\.30,17)\.30,19)\.30,21)\.30,23)\.30,25)\.30,27)\.31,14)
\.32,11)\.32,12)\.32,13)\.32,14)\.32,17)\.32,20)\.33,9)
\.33,10)\.33,11)\.33,14)\.33,15)\.33,16)\.33,20)\.33,24)
\.33,29)\.34,15)\.35,11)\.35,13)\.35,14)\.35,15)\.35,16)
\.35,17)\.35,18)\.35,19)\.35,20)\.35,23)\.35,25)\.35,26)
\.35,27)\.35,29)\.36,8)\.36,10)\.36,11)\.36,12)\.36,13)
\.36,14)\.36,15)\.36,16)\.36,17)\.36,18)\.36,19)\.36,20)
\.36,22)\.36,23)\.36,24)\.36,26)\.36,27)\.37,17)\.38,17)
\.39,11)\.39,12)\.39,13)\.39,14)\.39,15)\.39,17)\.39,18)
\.39,19)\.39,20)\.39,21)\.39,23)\.39,24)\.39,27)\.39,29)
\.40,11)\.40,13)\.40,14)\.40,15)\.40,16)\.40,17)\.40,18)
\.40,19)\.40,20)\.40,21)\.40,22)\.40,23)\.40,24)\.40,26)
\.40,27)\.40,31)\.40,32)\.40,33)\.41,19)\.42,12)\.42,13)
\.42,14)\.42,16)\.42,17)\.42,18)\.42,19)\.42,20)\.42,23)
\.42,25)\.42,26)\.42,29)\.42,31)\.42,33)\.42,35)\.43,20)
\.44,14)\.44,15)\.44,16)\.44,19)\.44,20)\.44,21)\.44,24)
\.44,29)\.44,30)\.44,31)\.44,34)\.45,11)\.45,13)\.45,14)
\.45,15)\.45,16)\.45,17)\.45,18)\.45,19)\.45,20)\.45,21)
\.45,22)\.45,23)\.45,24)\.45,25)\.45,26)\.45,27)\.45,29)
\.45,33)\.45,34)\.45,35)\.45,36)\.45,39)\.46,21)\.47,22)
\.48,11)\.48,13)\.48,14)\.48,15)\.48,16)\.48,17)\.48,18)
\.48,19)\.48,20)\.48,21)\.48,22)\.48,23)\.48,24)\.48,26)
\.48,27)\.48,29)\.48,31)\.48,32)\.48,33)\.48,34)\.49,20)
\.49,23)\.49,26)\.49,29)\.49,38)\.50,19)\.50,21)\.50,23)
\.50,27)\.50,35)\.50,39)\.51,15)\.51,18)\.51,23)\.51,25)
\.51,27)\.51,31)\.51,34)\.51,35)\.51,39)\.52,17)\.52,18)
\.52,20)\.52,23)\.52,24)\.52,25)\.52,26)\.52,27)\.52,29)
\.52,36)\.52,38)\.52,39)\.53,25)\.54,17)\.54,18)\.54,21)
\.54,22)\.54,24)\.54,25)\.54,26)\.54,27)\.54,31)\.54,39)
\.54,41)\.55,19)\.55,21)\.55,23)\.55,24)\.55,26)\.55,28)
\.55,29)\.55,31)\.55,34)\.55,39)\.55,41)\.55,43)\.56,17)
\.56,19)\.56,20)\.56,22)\.56,23)\.56,24)\.56,25)\.56,26)
\.56,28)\.56,29)\.56,32)\.56,35)\.56,38)\.56,43)\.57,17)
\.57,18)\.57,19)\.57,20)\.57,26)\.57,27)\.57,28)\.57,29)
\.57,32)\.57,35)\.57,36)\.58,27)\.59,28)\.60,11)\.60,14)
\.60,15)\.60,16)\.60,17)\.60,18)\.60,19)\.60,20)\.60,21)
\.60,22)\.60,23)\.60,24)\.60,25)\.60,26)\.60,27)\.60,28)
\.60,29)\.60,30)\.60,31)\.60,32)\.60,33)\.60,34)\.60,35)
\.60,36)\.60,37)\.60,38)\.60,39)\.60,41)\.60,44)\.61,29)
\.62,29)\.63,17)\.63,20)\.63,21)\.63,22)\.63,23)\.63,26)
\.63,27)\.63,28)\.63,29)\.63,32)\.63,33)\.63,34)\.63,35)
\.63,36)\.63,38)\.63,41)\.63,42)\.63,44)\.64,24)\.64,25)
\.64,26)\.64,27)\.64,28)\.64,29)\.64,32)\.64,34)\.64,35)
\.64,41)\.65,23)\.65,25)\.65,27)\.65,29)\.65,31)\.65,33)
\.65,35)\.65,37)\.65,41)\.65,43)\.65,47)\.65,49)\.65,51)
\.66,20)\.66,21)\.66,23)\.66,24)\.66,26)\.66,29)\.66,30)
\.66,31)\.66,34)\.66,39)\.66,41)\.66,44)\.66,49)\.67,32)
\.68,23)\.68,24)\.68,25)\.68,31)\.68,32)\.68,33)\.68,35)
\.68,39)\.68,49)\.68,50)\.69,21)\.69,22)\.69,32)\.69,33)
\.69,34)\.69,43)\.70,26)\.70,27)\.70,29)\.70,30)\.70,31)
\.70,33)\.70,35)\.70,39)\.70,43)\.70,47)\.70,49)\.70,51)
\.71,34)\.72,19)\.72,20)\.72,22)\.72,23)\.72,24)\.72,25)
\.72,26)\.72,27)\.72,28)\.72,29)\.72,30)\.72,31)\.72,32)
\.72,33)\.72,34)\.72,35)\.72,37)\.72,38)\.72,39)\.72,41)
\.72,43)\.72,44)\.72,46)\.72,47)\.73,35)\.74,35)\.75,19)
\.75,20)\.75,23)\.75,24)\.75,26)\.75,27)\.75,28)\.75,29)
\.75,30)\.75,31)\.75,33)\.75,34)\.75,35)\.75,36)\.75,38)
\.75,39)\.75,40)\.75,41)\.75,44)\.75,47)\.75,49)\.75,51)
\.76,26)\.76,29)\.76,35)\.76,36)\.76,37)\.76,44)\.76,54)
\.76,55)\.77,29)\.77,32)\.77,34)\.77,35)\.77,37)\.77,40)
\.77,47)\.77,49)\.77,50)\.78,25)\.78,27)\.78,29)\.78,31)
\.78,35)\.78,36)\.78,37)\.78,38)\.78,39)\.78,41)\.79,38)
\.80,23)\.80,25)\.80,27)\.80,28)\.80,29)\.80,30)\.80,31)
\.80,33)\.80,34)\.80,35)\.80,36)\.80,37)\.80,38)\.80,39)
\.80,41)\.80,42)\.80,43)\.80,44)\.80,47)\.80,48)\.80,49)
\.80,51)\.80,57)\.81,26)\.81,27)\.81,28)\.81,29)\.81,32)
\.81,34)\.81,35)\.81,36)\.81,37)\.81,38)\.81,39)\.81,41)
\.81,43)\.81,44)\.81,50)\.81,51)\.82,39)\.83,40)\.84,19)
\.84,21)\.84,23)\.84,25)\.84,26)\.84,27)\.84,28)\.84,29)
\.84,30)\.84,31)\.84,32)\.84,33)\.84,34)\.84,35)\.84,37)
\.84,38)\.84,39)\.84,40)\.84,41)\.84,43)\.84,44)\.84,46)
\.84,47)\.84,48)\.84,50)\.84,51)\.85,33)\.85,37)\.85,39)
\.85,41)\.85,43)\.85,45)\.85,47)\.85,49)\.85,65)\.86,41)
\.87,27)\.87,28)\.87,41)\.87,42)\.87,43)\.87,45)\.87,48)
\.87,56)\.87,57)\.88,31)\.88,32)\.88,36)\.88,37)\.88,39)
\.88,41)\.88,44)\.88,45)\.88,46)\.88,47)\.88,48)\.88,49)
\.88,50)\.89,43)\.90,25)\.90,26)\.90,29)\.90,30)\.90,31)
\.90,33)\.90,35)\.90,37)\.90,39)\.90,42)\.90,43)\.90,44)
\.90,47)\.90,48)\.90,49)\.90,55)\.90,65)\.91,35)\.91,38)
\.91,41)\.91,44)\.91,50)\.91,56)\.91,59)\.92,43)\.92,45)
\.92,54)\.93,29)\.93,30)\.93,32)\.93,39)\.93,44)\.93,46)
\.93,47)\.93,49)\.93,54)\.93,60)\.93,61)\.93,65)\.94,45)
\.95,35)\.95,37)\.95,39)\.95,44)\.95,46)\.95,48)\.95,50)
\.95,53)\.95,55)\.95,57)\.95,59)\.95,71)\.96,24)\.96,26)
\.96,28)\.96,32)\.96,34)\.96,35)\.96,36)\.96,37)\.96,38)
\.96,39)\.96,41)\.96,42)\.96,43)\.96,44)\.96,45)\.96,46)
\.96,47)\.96,49)\.96,50)\.96,53)\.96,55)\.97,47)\.98,44)
\.98,47)\.99,29)\.99,32)\.99,33)\.99,34)\.99,35)\.99,37)
\.99,38)\.99,42)\.99,45)\.99,46)\.99,47)\.99,48)\.99,49)
\.99,50)\.99,52)\.99,54)\.99,59)\.99,60)\.99,64)\.100,29)
\.100,31)\.100,33)\.100,37)\.100,39)\.100,41)\.100,42)\.100,44)
\.100,46)\.100,47)\.100,49)\.100,51)\.100,56)\.100,57)\.100,60)
\.101,49)\.102,32)\.102,33)\.102,47)\.102,49)\.102,50)\.102,65)
\.103,50)\.104,43)\.104,44)\.104,45)\.104,51)\.104,53)\.104,56)
\.105,23)\.105,27)\.105,29)\.105,33)\.105,34)\.105,35)\.105,36)
\.105,37)\.105,38)\.105,39)\.105,41)\.105,42)\.105,43)\.105,44)
\.105,45)\.105,46)\.105,47)\.105,48)\.105,49)\.105,50)\.105,51)
\.105,53)\.105,54)\.105,55)\.105,56)\.105,57)\.105,59)\.105,60)
\.105,62)\.105,65)\.105,68)\.105,69)\.106,51)\.107,52)\.108,32)
\.108,37)\.108,38)\.108,39)\.108,40)\.108,41)\.108,43)\.108,44)
\.108,45)\.108,46)\.108,47)\.108,49)\.108,50)\.108,52)\.108,53)
\.108,54)\.108,55)\.108,56)\.108,57)\.108,59)\.108,65)\.108,67)
\.109,53)\.110,43)\.110,44)\.110,49)\.110,51)\.110,53)\.110,57)
\.110,59)\.110,63)\.110,67)\.111,35)\.111,36)\.111,53)\.111,56)
\.111,72)\.111,74)\.111,75)\.112,40)\.112,41)\.112,42)\.112,43)
\.112,44)\.112,45)\.112,46)\.112,47)\.112,49)\.112,50)\.112,51)
\.112,53)\.112,58)\.112,59)\.112,63)\.112,64)\.112,65)\.112,68)
\.113,55)\.114,36)\.114,37)\.114,44)\.114,46)\.114,54)\.114,55)
\.115,43)\.115,54)\.115,56)\.115,65)\.115,67)\.115,69)\.116,41)
\.116,42)\.116,57)\.116,62)\.117,35)\.117,38)\.117,39)\.117,41)
\.117,44)\.117,50)\.117,51)\.117,53)\.117,56)\.117,59)\.117,62)
\.117,63)\.117,65)\.117,71)\.119,50)\.119,55)\.119,56)\.119,58)
\.119,61)\.119,66)\.119,69)\.119,71)\.119,74)\.119,77)\.120,31)
\.120,32)\.120,33)\.120,34)\.120,35)\.120,36)\.120,37)\.120,38)
\.120,39)\.120,41)\.120,42)\.120,43)\.120,44)\.120,45)\.120,46)
\.120,47)\.120,48)\.120,49)\.120,50)\.120,51)\.120,52)\.120,53)
\.120,54)\.120,55)\.120,56)\.120,57)\.120,58)\.120,59)\.120,60)
\.120,62)\.120,64)\.120,65)\.120,66)\.120,67)\.120,69)\.120,70)
\.120,72)\.120,75)\.121,59)\.121,64)\.122,59)\.123,39)\.123,40)
\.123,59)\.123,61)\.123,64)\.123,80)\.124,45)\.124,49)\.124,61)
\.124,65)\.125,49)\.125,51)\.125,59)\.125,61)\.125,63)\.125,65)
\.125,67)\.125,73)\.126,35)\.126,38)\.126,43)\.126,44)\.126,45)
\.126,47)\.126,48)\.126,49)\.126,52)\.126,53)\.126,54)\.126,55)
\.126,58)\.126,59)\.126,63)\.126,65)\.126,67)\.126,76)\.127,62)
\.128,47)\.128,49)\.128,51)\.128,53)\.128,54)\.128,64)\.128,68)
\.129,41)\.129,42)\.129,45)\.129,48)\.129,62)\.129,63)\.129,64)
\.129,85)\.130,49)\.130,51)\.130,59)\.130,63)\.130,71)\.131,64)
\.132,29)\.132,32)\.132,39)\.132,41)\.132,42)\.132,45)\.132,46)
\.132,48)\.132,50)\.132,52)\.132,53)\.132,54)\.132,55)\.132,59)
\.132,63)\.132,65)\.132,66)\.133,53)\.133,56)\.133,65)\.133,68)
\.133,71)\.133,74)\.133,77)\.133,80)\.133,83)\.134,65)\.135,37)
\.135,44)\.135,45)\.135,46)\.135,47)\.135,48)\.135,49)\.135,50)
\.135,53)\.135,54)\.135,55)\.135,56)\.135,57)\.135,59)\.135,60)
\.135,62)\.135,63)\.135,64)\.135,65)\.135,67)\.135,68)\.135,69)
\.135,71)\.135,74)\.135,77)\.135,86)\.136,50)\.136,53)\.136,57)
\.136,58)\.136,60)\.136,63)\.136,65)\.136,67)\.136,69)\.136,71)
\.137,67)\.138,45)\.138,54)\.138,56)\.138,89)\.139,68)\.140,39)
\.140,41)\.140,43)\.140,44)\.140,45)\.140,47)\.140,50)\.140,51)
\.140,53)\.140,54)\.140,57)\.140,58)\.140,59)\.140,62)\.140,63)
\.140,65)\.140,66)\.140,67)\.140,69)\.140,72)\.140,73)\.140,83)
\.140,85)\.141,45)\.141,68)\.141,69)\.142,69)\.143,65)\.143,70)
\.143,75)\.143,76)\.143,81)\.144,41)\.144,44)\.144,51)\.144,52)
\.144,55)\.144,56)\.144,57)\.144,59)\.144,60)\.144,62)\.144,63)
\.144,65)\.144,66)\.144,67)\.144,68)\.144,69)\.144,70)\.144,71)
\.144,74)\.144,79)\.144,82)\.144,85)\.144,86)\.145,55)\.145,57)
\.145,69)\.145,71)\.145,73)\.145,83)\.145,85)\.145,87)\.146,71)
\.147,41)\.147,47)\.147,48)\.147,50)\.147,51)\.147,53)\.147,55)
\.147,68)\.147,71)\.147,80)\.147,90)\.147,92)\.147,96)\.147,97)
\.148,56)\.148,71)\.148,75)\.148,77)\.149,73)\.150,47)\.150,49)
\.150,51)\.150,53)\.150,55)\.150,57)\.150,60)\.150,61)\.150,67)
\.150,69)\.150,74)\.151,74)\.152,55)\.152,56)\.152,65)\.152,66)
\.152,68)\.152,71)\.152,73)\.153,47)\.153,50)\.153,51)\.153,53)
\.153,55)\.153,59)\.153,63)\.153,64)\.153,66)\.153,69)\.153,72)
\.153,73)\.153,74)\.153,79)\.153,83)\.153,87)\.154,65)\.154,69)
\.154,70)\.154,75)\.154,79)\.154,81)\.154,85)\.155,59)\.155,61) }{
\.155,74)\.155,76)\.155,78)\.155,80)\.155,86)\.155,91)\.155,93)
\.155,95)\.156,37)\.156,51)\.156,52)\.156,53)\.156,56)\.156,57)
\.156,59)\.156,61)\.156,63)\.156,65)\.156,67)\.156,71)\.156,75)
\.156,76)\.156,78)\.156,99)\.157,77)\.159,51)\.159,52)\.159,78)
\.159,79)\.160,53)\.160,55)\.160,60)\.160,63)\.160,65)\.160,66)
\.160,67)\.160,68)\.160,69)\.160,70)\.160,71)\.160,72)\.160,76)
\.160,79)\.160,89)\.160,93)\.161,65)\.161,68)\.161,71)\.161,76)
\.161,79)\.161,82)\.161,101)\.162,53)\.162,61)\.162,66)\.162,67)
\.162,69)\.163,80)\.164,59)\.164,61)\.164,63)\.164,64)\.165,39)
\.165,47)\.165,53)\.165,54)\.165,56)\.165,57)\.165,58)\.165,59)
\.165,60)\.165,63)\.165,64)\.165,65)\.165,66)\.165,67)\.165,68)
\.165,69)\.165,70)\.165,71)\.165,75)\.165,76)\.165,84)\.165,86)
\.165,88)\.165,91)\.165,93)\.165,109)\.166,81)\.167,82)\.168,40)
\.168,43)\.168,46)\.168,48)\.168,49)\.168,50)\.168,53)\.168,55)
\.168,57)\.168,58)\.168,61)\.168,62)\.168,63)\.168,64)\.168,65)
\.168,66)\.168,67)\.168,68)\.168,70)\.168,71)\.168,75)\.168,76)
\.168,77)\.168,80)\.168,83)\.168,84)\.168,85)\.168,86)\.168,87)
\.168,90)\.168,101)\.168,103)\.169,77)\.169,83)\.169,89)\.169,95)
\.170,67)\.170,71)\.170,75)\.170,79)\.170,81)\.170,83)\.171,65)
\.171,71)\.171,74)\.171,77)\.171,78)\.171,86)\.171,91)\.171,99)
\.171,101)\.171,110)\.172,62)\.172,63)\.173,85)\.174,59)\.174,62)
\.174,69)\.174,71)\.175,69)\.175,71)\.175,72)\.175,73)\.175,74)
\.175,76)\.175,77)\.175,79)\.175,84)\.175,86)\.175,88)\.175,92)
\.175,94)\.175,95)\.175,98)\.176,63)\.176,65)\.176,69)\.176,70)
\.176,71)\.176,73)\.176,76)\.176,78)\.176,80)\.176,85)\.176,91)
\.176,93)\.176,96)\.177,57)\.177,86)\.177,87)\.178,87)\.179,88)
\.180,47)\.180,53)\.180,54)\.180,55)\.180,56)\.180,57)\.180,59)
\.180,61)\.180,62)\.180,63)\.180,64)\.180,65)\.180,66)\.180,67)
\.180,68)\.180,69)\.180,70)\.180,71)\.180,72)\.180,73)\.180,74)
\.180,75)\.180,76)\.180,77)\.180,78)\.180,79)\.180,80)\.180,83)
\.180,86)\.180,92)\.180,98)\.180,99)\.180,110)\.181,89)\.182,77)
\.182,83)\.183,60)\.183,61)\.183,89)\.183,92)\.183,95)\.183,120)
\.184,76)\.184,81)\.184,82)\.184,89)\.185,71)\.185,73)\.185,95)
\.185,97)\.185,109)\.186,61)\.186,65)\.186,76)\.186,78)\.186,86)
\.187,79)\.187,84)\.187,87)\.187,92)\.187,97)\.187,100)\.187,105)
\.188,68)\.188,91)\.188,92)\.189,53)\.189,65)\.189,66)\.189,68)
\.189,70)\.189,71)\.189,72)\.189,74)\.189,82)\.189,83)\.189,85)
\.189,86)\.189,90)\.189,91)\.189,95)\.189,98)\.189,101)\.189,104)
\.189,116)\.190,75)\.190,89)\.190,107)\.191,94)\.192,47)\.192,54)
\.192,55)\.192,66)\.192,69)\.192,71)\.192,73)\.192,75)\.192,77)
\.192,79)\.192,80)\.192,82)\.192,83)\.192,84)\.192,85)\.192,86)
\.192,87)\.192,88)\.192,89)\.192,91)\.192,112)\.192,120)\.193,95)
\.195,51)\.195,63)\.195,66)\.195,67)\.195,70)\.195,71)\.195,75)
\.195,77)\.195,78)\.195,79)\.195,81)\.195,83)\.195,85)\.195,95)
\.195,100)\.195,103)\.195,113)\.196,65)\.196,71)\.196,72)\.196,74)
\.196,78)\.196,80)\.196,84)\.196,86)\.196,91)\.196,95)\.196,97)
\.196,99)\.196,101)\.196,105)\.197,97)\.198,67)\.198,79)\.198,82)
\.198,86)\.198,87)\.198,90)\.198,91)\.198,92)\.198,93)\.199,98)
\.200,69)\.200,72)\.200,73)\.200,74)\.200,76)\.200,78)\.200,80)
\.200,83)\.200,89)\.200,90)\.200,91)\.200,92)\.200,93)\.200,95)
\.200,97)\.200,98)\.200,105)\.201,65)\.201,66)\.201,68)\.202,99)
\.203,83)\.203,86)\.203,89)\.203,100)\.203,117)\.204,47)\.204,50)
\.204,53)\.204,68)\.204,69)\.204,71)\.204,74)\.204,75)\.204,81)
\.204,87)\.204,90)\.204,93)\.204,96)\.204,97)\.204,99)\.204,100)
\.205,81)\.205,89)\.205,101)\.205,103)\.205,123)\.206,101)\.207,65)
\.207,68)\.207,90)\.207,92)\.207,102)\.207,103)\.207,106)\.207,131)
\.208,71)\.208,77)\.208,81)\.208,82)\.208,83)\.208,88)\.208,89)
\.208,97)\.208,100)\.208,103)\.209,94)\.209,98)\.209,103)\.209,108)
\.209,112)\.209,117)\.210,62)\.210,64)\.210,65)\.210,69)\.210,70)
\.210,71)\.210,72)\.210,73)\.210,75)\.210,77)\.210,79)\.210,80)
\.210,84)\.210,85)\.210,86)\.210,87)\.210,90)\.210,94)\.210,96)
\.210,97)\.210,103)\.211,104)\.212,79)\.213,70)\.213,104)\.215,83)
\.215,87)\.215,106)\.215,125)\.216,56)\.216,64)\.216,68)\.216,74)
\.216,75)\.216,77)\.216,79)\.216,80)\.216,86)\.216,87)\.216,89)
\.216,91)\.216,95)\.216,97)\.216,98)\.216,99)\.216,100)\.216,101)
\.216,104)\.217,89)\.217,98)\.217,107)\.217,113)\.218,107)\.219,71)
\.219,72)\.219,115)\.220,63)\.220,65)\.220,66)\.220,71)\.220,79)
\.220,84)\.220,85)\.220,86)\.220,87)\.220,91)\.220,102)\.220,103)
\.220,104)\.220,109)\.220,125)\.221,95)\.221,101)\.221,103)\.221,109)
\.221,115)\.221,117)\.222,75)\.222,89)\.222,91)\.222,97)\.223,110)
\.224,74)\.224,80)\.224,85)\.224,86)\.224,87)\.224,89)\.224,90)
\.224,95)\.224,97)\.224,101)\.224,102)\.224,104)\.224,105)\.225,59)
\.225,77)\.225,78)\.225,80)\.225,81)\.225,83)\.225,86)\.225,89)
\.225,92)\.225,95)\.225,96)\.225,98)\.225,99)\.225,101)\.225,104)
\.225,111)\.225,116)\.225,122)\.225,123)\.226,111)\.227,112)\.228,55)
\.228,56)\.228,86)\.228,89)\.228,91)\.228,94)\.228,102)\.228,105)
\.228,110)\.229,113)\.230,91)\.231,65)\.231,66)\.231,69)\.231,75)
\.231,79)\.231,81)\.231,85)\.231,86)\.231,87)\.231,89)\.231,101)
\.231,107)\.231,108)\.231,109)\.231,114)\.231,118)\.231,122)\.232,100)
\.232,101)\.232,115)\.232,117)\.233,115)\.234,83)\.234,89)\.234,95)
\.234,98)\.234,101)\.234,103)\.234,109)\.235,91)\.235,93)\.235,95)
\.235,116)\.236,86)\.236,115)\.237,77)\.237,78)\.237,81)\.237,116)
\.238,107)\.238,117)\.239,118)\.240,47)\.240,55)\.240,60)\.240,67)
\.240,68)\.240,70)\.240,71)\.240,72)\.240,75)\.240,76)\.240,83)
\.240,84)\.240,85)\.240,86)\.240,87)\.240,89)\.240,93)\.240,97)
\.240,99)\.240,101)\.240,104)\.240,107)\.240,109)\.240,110)\.240,112)
\.240,115)\.240,117)\.240,118)\.240,120)\.240,126)\.240,127)\.240,135)
\.241,119)\.242,114)\.242,129)\.243,92)\.243,106)\.243,111)\.243,118)
\.243,122)\.243,124)\.244,90)\.244,95)\.245,95)\.245,97)\.245,103)
\.245,106)\.245,124)\.245,125)\.246,99)\.247,107)\.247,116)\.247,122)
\.247,128)\.248,92)\.248,113)\.248,121)\.249,81)\.249,82)\.249,124)
\.250,99)\.250,123)\.251,124)\.252,53)\.252,64)\.252,68)\.252,80)
\.252,83)\.252,86)\.252,89)\.252,91)\.252,92)\.252,93)\.252,95)
\.252,101)\.252,106)\.252,108)\.252,110)\.252,111)\.252,114)\.252,116)
\.252,118)\.252,121)\.252,122)\.253,125)\.253,130)\.253,131)\.253,136)
\.253,141)\.255,81)\.255,83)\.255,86)\.255,90)\.255,92)\.255,95)
\.255,97)\.255,99)\.255,102)\.255,103)\.255,109)\.255,113)\.255,115)
\.255,120)\.255,121)\.255,123)\.255,131)\.255,145)\.256,97)\.256,101)
\.256,102)\.256,103)\.256,110)\.256,112)\.256,117)\.256,118)\.256,120)
\.256,121)\.256,122)\.257,127)\.258,106)\.258,126)\.259,113)\.259,116)
\.259,128)\.260,77)\.260,83)\.260,84)\.260,89)\.260,95)\.260,97)
\.260,98)\.260,100)\.260,103)\.260,104)\.260,121)\.260,122)\.260,125)
\.261,86)\.261,88)\.261,89)\.261,111)\.261,114)\.261,115)\.261,117)
\.261,119)\.261,126)\.263,130)\.264,69)\.264,76)\.264,78)\.264,79)
\.264,81)\.264,85)\.264,89)\.264,92)\.264,93)\.264,96)\.264,101)
\.264,107)\.264,113)\.264,114)\.264,121)\.264,122)\.264,123)\.264,126)
\.264,136)\.265,103)\.265,105)\.265,129)\.265,131)\.265,133)\.266,125)
\.266,131)\.267,87)\.267,88)\.267,133)\.268,98)\.268,133)\.269,133)
\.270,80)\.270,89)\.270,91)\.270,95)\.270,99)\.270,101)\.270,105)
\.270,108)\.270,109)\.270,112)\.270,126)\.270,127)\.270,128)\.271,134)
\.272,100)\.272,103)\.272,107)\.272,109)\.272,115)\.272,116)\.272,119)
\.272,132)\.273,77)\.273,80)\.273,83)\.273,85)\.273,89)\.273,95)
\.273,96)\.273,97)\.273,104)\.273,105)\.273,109)\.273,118)\.273,119)
\.273,126)\.273,129)\.273,131)\.273,145)\.274,135)\.275,109)\.275,134)
\.275,136)\.275,138)\.275,143)\.275,145)\.275,146)\.275,148)\.276,69)
\.276,89)\.276,98)\.276,99)\.276,103)\.276,109)\.276,130)\.276,131)
\.279,123)\.279,137)\.280,83)\.280,103)\.280,105)\.280,106)\.280,115)
\.280,121)\.280,124)\.280,125)\.280,126)\.280,130)\.280,131)\.280,143)
\.280,151)\.282,93)\.283,140)\.284,104)\.285,75)\.285,94)\.285,95)
\.285,102)\.285,104)\.285,107)\.285,108)\.285,113)\.285,117)\.285,122)
\.285,128)\.285,142)\.286,125)\.287,119)\.287,122)\.287,142)\.287,145)
\.288,86)\.288,89)\.288,100)\.288,101)\.288,112)\.288,118)\.288,122)
\.288,123)\.288,126)\.288,130)\.288,131)\.288,137)\.288,141)\.288,143)
\.289,143)\.289,151)\.290,125)\.290,129)\.290,147)\.291,96)\.291,146)
\.292,108)\.292,147)\.293,145)\.294,86)\.294,91)\.294,97)\.294,101)
\.294,103)\.294,105)\.294,122)\.294,127)\.295,115)\.295,146)\.296,128)
\.296,147)\.297,89)\.297,94)\.297,98)\.297,101)\.297,102)\.297,105)
\.297,124)\.297,130)\.297,132)\.297,135)\.297,138)\.297,144)\.299,154)
\.300,74)\.300,80)\.300,91)\.300,93)\.300,98)\.300,103)\.300,104)
\.300,105)\.300,114)\.300,124)\.300,128)\.300,136)\.300,141)\.300,142)
\.300,144)\.300,146)\.301,131)\.301,149)\.303,99)\.303,100)\.303,151)
\.304,116)\.304,121)\.304,123)\.304,125)\.304,130)\.304,139)\.304,141)
\.304,157)\.305,119)\.305,121)\.305,127)\.305,151)\.306,95)\.306,111)
\.306,113)\.306,130)\.307,152)\.308,89)\.308,101)\.308,102)\.308,105)
\.308,132)\.308,146)\.308,160)\.309,102)\.309,103)\.311,154)\.312,71)
\.312,88)\.312,96)\.312,97)\.312,100)\.312,104)\.312,109)\.312,133)
\.312,134)\.312,143)\.312,147)\.312,151)\.312,152)\.313,155)\.315,87)
\.315,89)\.315,95)\.315,110)\.315,111)\.315,114)\.315,119)\.315,124)
\.315,131)\.315,133)\.315,135)\.315,137)\.315,140)\.315,144)\.315,155)
\.316,116)\.316,156)\.318,129)\.318,131)\.319,144)\.319,153)\.319,167)
\.319,172)\.320,97)\.320,101)\.320,117)\.320,120)\.320,121)\.320,123)
\.320,127)\.320,140)\.320,143)\.320,144)\.320,148)\.320,150)\.320,151)
\.320,153)\.320,158)\.321,105)\.322,143)\.322,153)\.322,165)\.323,152)
\.323,160)\.324,116)\.324,118)\.324,120)\.324,124)\.324,142)\.324,152)
\.324,154)\.324,155)\.325,129)\.325,131)\.325,143)\.325,163)\.325,165)
\.325,167)\.325,173)\.327,108)\.328,144)\.328,145)\.328,165)\.329,140)
\.329,163)\.329,166)\.330,98)\.330,104)\.330,109)\.330,114)\.330,118)
\.330,125)\.330,128)\.330,129)\.330,138)\.331,164)\.333,110)\.333,113) }{
\.333,155)\.336,86)\.336,97)\.336,101)\.336,102)\.336,109)\.336,110)
\.336,114)\.336,119)\.336,120)\.336,122)\.336,123)\.336,127)\.336,129)
\.336,131)\.336,135)\.336,140)\.336,141)\.336,147)\.336,148)\.336,151)
\.336,153)\.336,155)\.337,167)\.338,161)\.339,111)\.340,102)\.340,103)
\.340,120)\.340,126)\.340,127)\.340,134)\.340,137)\.340,152)\.340,161)
\.341,164)\.341,174)\.341,179)\.342,113)\.342,116)\.342,117)\.342,122)
\.342,143)\.342,150)\.342,157)\.343,149)\.343,176)\.344,149)\.345,116)
\.345,129)\.345,130)\.345,135)\.345,141)\.345,148)\.345,154)\.345,166)
\.348,117)\.349,173)\.350,149)\.351,122)\.351,161)\.351,165)\.351,168)
\.351,172)\.351,173)\.352,141)\.352,144)\.352,154)\.352,164)\.352,168)
\.352,169)\.354,117)\.356,132)\.356,133)\.356,175)\.357,101)\.357,107)
\.357,111)\.357,117)\.357,118)\.357,121)\.357,123)\.357,125)\.357,126)
\.357,170)\.360,106)\.360,117)\.360,122)\.360,123)\.360,125)\.360,126)
\.360,129)\.360,131)\.360,141)\.360,145)\.360,148)\.360,150)\.360,154)
\.360,157)\.360,158)\.360,161)\.360,164)\.360,173)\.360,174)\.360,176)
\.360,178)\.361,179)\.363,119)\.363,123)\.363,129)\.363,165)\.363,184)
\.364,116)\.364,132)\.364,164)\.364,180)\.365,145)\.365,181)\.366,123)
\.368,131)\.368,147)\.369,122)\.369,187)\.370,147)\.371,158)\.371,184)
\.372,92)\.372,123)\.373,185)\.375,103)\.375,119)\.375,123)\.375,126)
\.375,127)\.375,161)\.376,166)\.376,185)\.377,173)\.377,181)\.378,113)
\.378,128)\.378,131)\.378,173)\.378,180)\.379,188)\.380,114)\.380,117)
\.380,143)\.381,125)\.381,126)\.383,190)\.384,110)\.384,112)\.384,118)
\.384,120)\.384,122)\.384,175)\.384,183)\.385,151)\.385,186)\.387,193)
\.388,146)\.389,193)\.390,122)\.390,124)\.390,129)\.390,156)\.391,186)
\.391,194)\.392,149)\.392,170)\.392,173)\.392,176)\.393,129)\.393,130)
\.395,157)\.395,200)\.396,89)\.396,101)\.396,125)\.396,129)\.396,133)
\.396,134)\.396,146)\.396,165)\.396,172)\.396,173)\.396,185)\.396,186)
\.396,190)\.399,115)\.399,116)\.399,134)\.399,143)\.399,175)\.399,190)
\.399,198)\.400,157)\.400,167)\.400,181)\.401,199)\.402,133)\.403,194)
\.403,206)\.404,151)\.405,138)\.405,140)\.405,145)\.405,147)\.405,163)
\.405,174)\.405,176)\.405,195)\.405,196)\.408,100)\.408,116)\.408,117)
\.408,118)\.408,121)\.408,137)\.408,139)\.408,170)\.408,185)\.408,186)
\.411,135)\.413,179)\.414,143)\.414,152)\.414,164)\.414,183)\.415,204)
\.415,206)\.416,157)\.416,168)\.416,180)\.416,194)\.416,200)\.416,201)
\.417,137)\.417,138)\.417,139)\.418,207)\.420,103)\.420,119)\.420,121)
\.420,124)\.420,125)\.420,126)\.420,130)\.420,131)\.420,134)\.420,143)
\.420,144)\.420,146)\.420,151)\.420,152)\.420,156)\.420,163)\.420,184)
\.420,190)\.420,194)\.420,198)\.420,202)\.420,204)\.420,208)\.423,142)
\.427,182)\.429,129)\.429,143)\.429,151)\.432,134)\.432,138)\.432,147)
\.432,155)\.432,168)\.432,180)\.432,185)\.432,187)\.432,214)\.434,185)
\.435,147)\.435,148)\.435,150)\.435,152)\.435,158)\.435,159)\.435,167)
\.435,210)\.438,145)\.438,147)\.439,218)\.440,191)\.440,198)\.440,217)
\.441,146)\.441,149)\.441,155)\.441,212)\.445,175)\.445,177)\.448,199)
\.448,208)\.448,214)\.448,218)\.448,222)\.450,134)\.450,145)\.450,218)
\.451,229)\.453,151)\.455,183)\.455,194)\.455,197)\.455,220)\.455,229)
\.456,130)\.456,133)\.456,134)\.456,139)\.456,157)\.456,196)\.459,143)
\.459,152)\.460,140)\.460,141)\.460,164)\.460,184)\.461,229)\.462,129)
\.462,133)\.462,147)\.462,158)\.464,195)\.464,204)\.465,149)\.465,153)
\.465,154)\.465,158)\.465,164)\.465,169)\.465,174)\.465,194)\.465,200)
\.465,206)\.468,165)\.468,220)\.468,230)\.471,158)\.473,251)\.476,206)
\.480,120)\.480,139)\.480,140)\.480,148)\.480,150)\.480,151)\.480,158)
\.480,161)\.480,165)\.480,169)\.480,173)\.480,200)\.480,204)\.480,208)
\.480,227)\.480,229)\.481,245)\.483,137)\.483,143)\.483,153)\.483,159)
\.483,165)\.483,212)\.484,165)\.486,214)\.488,212)\.489,165)\.492,165)
\.492,166)\.493,245)\.495,180)\.495,213)\.495,214)\.495,231)\.495,246)
\.496,194)\.496,231)\.500,149)\.504,143)\.504,148)\.504,149)\.504,173)
\.504,176)\.504,179)\.504,181)\.504,212)\.504,214)\.504,218)\.504,242)
\.505,201)\.507,167)\.507,168)\.507,173)\.507,175)\.507,180)\.510,143)
\.510,155)\.510,161)\.511,218)\.511,254)\.513,178)\.513,228)\.516,193)
\.520,157)\.520,201)\.520,229)\.520,250)\.522,167)\.522,173)\.522,181)
\.525,149)\.525,151)\.525,155)\.525,167)\.525,177)\.525,181)\.525,190)
\.525,210)\.525,225)\.525,230)\.528,154)\.528,164)\.528,168)\.528,231)
\.528,238)\.528,243)\.531,173)\.531,176)\.532,228)\.537,178)\.540,163)
\.540,178)\.540,182)\.540,200)\.540,210)\.540,251)\.543,180)\.544,238)
\.544,264)\.546,155)\.546,164)\.546,173)\.546,174)\.546,180)\.546,241)
\.546,264)\.552,131)\.552,163)\.552,165)\.552,185)\.552,243)\.555,183)
\.555,185)\.555,202)\.555,239)\.555,245)\.560,241)\.560,242)\.564,185)
\.564,188)\.567,191)\.567,245)\.567,275)\.570,178)\.570,190)\.571,284)
\.576,167)\.576,271)\.579,192)\.579,194)\.580,173)\.580,210)\.585,196)
\.585,197)\.585,217)\.585,227)\.585,250)\.585,291)\.588,173)\.588,190)
\.591,196)\.595,260)\.597,198)\.600,167)\.600,174)\.600,175)\.600,181)
\.600,196)\.602,263)\.608,264)\.609,173)\.609,176)\.609,179)\.609,195)
\.609,254)\.612,272)\.615,229)\.615,259)\.615,301)\.616,257)\.616,301)
\.624,180)\.624,264)\.627,207)\.627,208)\.630,182)\.630,220)\.630,221)
\.630,263)\.630,271)\.632,275)\.645,230)\.648,190)\.648,218)\.648,221)
\.648,275)\.651,185)\.651,188)\.651,218)\.651,222)\.651,280)\.651,282)
\.657,218)\.660,198)\.660,214)\.660,318)\.669,223)\.672,195)\.672,199)
\.672,208)\.672,214)\.672,218)\.672,222)\.672,287)\.672,321)\.675,228)
\.675,231)\.675,246)\.675,291)\.675,330)\.676,330)\.687,228)\.693,214)
\.693,236)\.696,195)\.696,204)\.696,205)\.700,210)\.700,306)\.705,229)
\.707,302)\.714,206)\.714,212)\.717,238)\.720,231)\.723,240)\.744,218)
\.744,222)\.752,321)\.756,221)\.756,260)\.756,320)\.759,258)\.765,330)
\.774,263)\.777,222)\.777,224)\.777,252)\.780,227)\.780,330)\.780,376)
\.783,251)\.791,341)\.792,257)\.795,264)\.795,265)\.798,228)\.798,259)
\.812,348)\.816,238)\.816,264)\.819,275)\.825,301)\.840,241)\.840,242)
\.840,275)\.840,301)\.840,355)\.867,272)\.867,291)\.870,251)\.885,295)
\.885,318)\.885,376)\.896,377)\.900,302)\.903,251)\.903,257)\.903,258)
\.903,263)\.903,306)\.903,387)\.912,264)\.924,257)\.924,302)\.936,275)
\.945,269)\.966,269)\.966,280)\.975,326)\.987,287)\.987,321)\.987,416)
\.1008,335)\.1029,293)\.1047,348)\.1050,302)\.1050,306)\.1113,320)\.1128,321)
\.1152,377)\.1155,335)\.1155,387)\.1218,348)\.1239,355)\.1299,433)\.1323,377)
\.1344,377)\.1407,462)\.1449,416)\.1617,462)\.1743,491)} \ve

\figc  \hocut{
\.4,0)\.6,0)\.8,1)\.10,1)\.12,1)\.12,2)\.12,3)
\.12,7)\.12,9)\.12,10)\.14,2)\.16,3)\.16,4)\.16,10)
\.18,2)\.18,3)\.18,4)\.18,8)\.18,10)\.20,3)\.20,4)
\.20,5)\.20,6)\.20,7)\.20,9)\.20,11)\.20,13)\.20,17)
\.20,18)\.20,19)\.20,23)\.22,4)\.24,3)\.24,4)\.24,5)
\.24,6)\.24,7)\.24,8)\.24,9)\.24,10)\.24,11)\.24,12)
\.24,13)\.24,16)\.24,23)\.26,5)\.28,5)\.28,8)\.28,9)
\.28,11)\.28,13)\.28,15)\.28,16)\.28,18)\.28,23)\.28,26)
\.30,3)\.30,4)\.30,5)\.30,6)\.30,7)\.30,8)\.30,9)
\.30,11)\.30,13)\.30,15)\.30,16)\.30,17)\.30,19)\.30,23)
\.32,7)\.32,8)\.32,9)\.32,10)\.32,14)\.34,7)\.36,5)
\.36,6)\.36,7)\.36,8)\.36,9)\.36,10)\.36,11)\.36,12)
\.36,13)\.36,14)\.36,15)\.36,16)\.36,17)\.36,19)\.36,20)
\.36,23)\.36,27)\.38,8)\.40,7)\.40,9)\.40,10)\.40,11)
\.40,12)\.40,13)\.40,14)\.40,15)\.40,17)\.40,19)\.40,21)
\.40,23)\.40,25)\.42,5)\.42,6)\.42,7)\.42,8)\.42,9)
\.42,10)\.42,11)\.42,12)\.42,13)\.42,15)\.42,16)\.42,17)
\.42,18)\.42,21)\.42,23)\.42,24)\.42,25)\.42,26)\.44,9)
\.44,10)\.44,11)\.44,14)\.44,16)\.44,20)\.44,23)\.46,10)
\.48,7)\.48,8)\.48,9)\.48,10)\.48,11)\.48,12)\.48,13)
\.48,14)\.48,15)\.48,16)\.48,17)\.48,18)\.48,19)\.48,20)
\.48,23)\.48,24)\.48,26)\.48,28)\.48,32)\.50,9)\.50,11)
\.50,13)\.50,17)\.50,25)\.52,11)\.52,13)\.52,14)\.52,17)
\.52,20)\.52,23)\.52,26)\.52,27)\.54,8)\.54,9)\.54,10)
\.54,11)\.54,12)\.54,14)\.54,15)\.54,16)\.54,19)\.54,20)
\.54,21)\.54,23)\.54,26)\.54,32)\.56,11)\.56,14)\.56,16)
\.56,17)\.56,18)\.56,19)\.56,20)\.56,23)\.56,26)\.56,29)
\.58,13)\.60,7)\.60,10)\.60,11)\.60,12)\.60,13)\.60,14)
\.60,15)\.60,16)\.60,17)\.60,18)\.60,19)\.60,20)\.60,21)
\.60,22)\.60,23)\.60,24)\.60,25)\.60,26)\.60,27)\.60,28)
\.60,29)\.60,30)\.60,31)\.60,33)\.60,35)\.62,14)\.64,15)
\.64,16)\.64,18)\.64,19)\.64,20)\.64,24)\.64,26)\.64,29)
\.66,9)\.66,10)\.66,14)\.66,19)\.66,20)\.66,23)\.66,29)
\.66,39)\.68,15)\.68,17)\.68,19)\.68,23)\.68,31)\.70,11)
\.70,13)\.70,14)\.70,16)\.70,17)\.70,18)\.70,23)\.70,25)
\.70,27)\.72,11)\.72,14)\.72,15)\.72,16)\.72,18)\.72,19)
\.72,20)\.72,21)\.72,22)\.72,23)\.72,24)\.72,25)\.72,26)
\.72,27)\.72,28)\.72,30)\.72,31)\.72,32)\.74,17)\.76,17)
\.76,20)\.76,23)\.76,26)\.78,11)\.78,12)\.78,13)\.78,17)
\.78,18)\.78,19)\.78,20)\.78,23)\.78,27)\.80,15)\.80,19)
\.80,20)\.80,21)\.80,23)\.80,24)\.80,25)\.80,26)\.80,27)
\.80,28)\.80,29)\.80,30)\.80,31)\.80,32)\.80,33)\.82,19)
\.84,11)\.84,13)\.84,14)\.84,15)\.84,16)\.84,17)\.84,19)
\.84,20)\.84,23)\.84,24)\.84,25)\.84,26)\.84,29)\.84,31)
\.84,34)\.84,35)\.84,37)\.84,38)\.84,40)\.84,41)\.84,44)
\.84,45)\.84,46)\.86,20)\.88,19)\.88,24)\.88,26)\.88,27)
\.88,34)\.90,11)\.90,14)\.90,15)\.90,16)\.90,17)\.90,18)
\.90,19)\.90,20)\.90,21)\.90,22)\.90,23)\.90,24)\.90,25)
\.90,26)\.90,27)\.90,28)\.90,29)\.90,30)\.90,34)\.90,35)
\.92,21)\.92,23)\.92,33)\.94,22)\.96,15)\.96,16)\.96,18)
\.96,20)\.96,22)\.96,23)\.96,24)\.96,25)\.96,26)\.96,28)
\.96,29)\.96,31)\.96,33)\.96,34)\.96,35)\.96,36)\.96,37)
\.96,39)\.96,41)\.96,43)\.96,45)\.98,20)\.98,23)\.98,26)
\.98,38)\.100,23)\.100,25)\.100,27)\.100,31)\.100,37)\.100,44)
\.100,49)\.102,15)\.102,17)\.102,23)\.102,24)\.102,27)\.102,35)
\.102,39)\.104,29)\.104,31)\.106,25)\.108,17)\.108,18)\.108,19)
\.108,21)\.108,22)\.108,23)\.108,25)\.108,27)\.108,29)\.108,30)
\.108,31)\.108,32)\.108,37)\.108,38)\.108,40)\.108,41)\.108,46)
\.108,50)\.108,56)\.110,21)\.110,24)\.110,26)\.110,28)\.110,34)
\.110,39)\.110,41)\.110,43)\.112,26)\.112,30)\.112,32)\.112,33)
\.112,34)\.112,35)\.112,36)\.112,37)\.112,38)\.112,40)\.112,41)
\.112,45)\.112,46)\.114,17)\.114,19)\.114,20)\.114,23)\.114,26)
\.114,27)\.114,36)\.116,27)\.116,49)\.118,28)\.120,15)\.120,19)
\.120,20)\.120,22)\.120,23)\.120,24)\.120,25)\.120,26)\.120,27)
\.120,28)\.120,29)\.120,30)\.120,31)\.120,32)\.120,33)\.120,34)
\.120,36)\.120,37)\.120,38)\.120,39)\.120,40)\.120,41)\.120,42)
\.120,43)\.120,44)\.120,46)\.120,47)\.120,48)\.120,49)\.120,51)
\.120,55)\.120,56)\.120,62)\.122,29)\.124,29)\.124,44)\.124,46)
\.126,17)\.126,20)\.126,21)\.126,22)\.126,23)\.126,25)\.126,26)
\.126,28)\.126,29)\.126,32)\.126,35)\.126,36)\.126,37)\.126,38)
\.126,41)\.126,44)\.126,45)\.128,32)\.128,36)\.128,40)\.128,41)
\.128,44)\.128,46)\.130,25)\.130,27)\.130,29)\.130,31)\.130,37)
\.130,43)\.130,47)\.132,19)\.132,22)\.132,23)\.132,29)\.132,34)
\.132,35)\.132,45)\.132,46)\.132,52)\.132,59)\.132,62)\.132,65)
\.132,67)\.134,32)\.136,33)\.136,37)\.136,41)\.136,50)\.136,53)
\.138,21)\.138,22)\.138,34)\.138,43)\.138,44)\.140,23)\.140,29)
\.140,32)\.140,39)\.140,41)\.140,43)\.140,47)\.140,53)\.140,54)
\.140,58)\.140,59)\.140,61)\.142,34)\.144,23)\.144,28)\.144,29)
\.144,30)\.144,35)\.144,37)\.144,38)\.144,40)\.144,41)\.144,43)
\.144,44)\.144,46)\.144,47)\.144,49)\.144,53)\.144,55)\.144,56)
\.144,62)\.146,35)\.148,35)\.148,57)\.150,19)\.150,23)\.150,24)
\.150,27)\.150,28)\.150,31)\.150,33)\.150,34)\.150,35)\.150,36)
\.150,39)\.150,49)\.150,51)\.150,55)\.152,44)\.152,46)\.152,48)
\.152,50)\.152,52)\.152,53)\.154,29)\.154,32)\.154,37)\.154,47)
\.156,23)\.156,25)\.156,27)\.156,28)\.156,52)\.156,56)\.156,61)
\.156,62)\.156,65)\.156,71)\.156,73)\.156,76)\.156,81)\.158,38)
\.160,33)\.160,39)\.160,41)\.160,44)\.160,45)\.160,46)\.160,48)
\.160,51)\.160,52)\.160,53)\.160,54)\.160,56)\.160,57)\.160,59)
\.160,62)\.162,26)\.162,32)\.162,33)\.162,34)\.162,41)\.162,44)
\.162,46)\.162,50)\.162,56)\.164,61)\.164,63)\.166,40)\.168,23)
\.168,28)\.168,30)\.168,31)\.168,32)\.168,33)\.168,34)\.168,35)
\.168,38)\.168,39)\.168,40)\.168,41)\.168,43)\.168,45)\.168,46)
\.168,47)\.168,49)\.168,52)\.168,55)\.168,56)\.168,57)\.168,61)
\.168,62)\.168,63)\.168,64)\.170,33)\.170,39)\.170,41)\.170,43)
\.170,47)\.170,49)\.172,43)\.172,44)\.172,62)\.174,27)\.174,28)
\.174,41)\.174,42)\.174,43)\.174,45)\.174,49)\.174,55)\.174,56)
\.176,58)\.176,60)\.176,64)\.178,43)\.180,29)\.180,31)\.180,32)
\.180,34)\.180,37)\.180,39)\.180,40)\.180,41)\.180,42)\.180,45)
\.180,46)\.180,49)\.180,50)\.180,52)\.180,53)\.180,54)\.180,55)
\.180,56)\.180,59)\.180,61)\.180,62)\.180,65)\.180,69)\.180,72)
\.180,73)\.180,79)\.180,86)\.182,35)\.182,44)\.182,47)\.182,56)
\.184,43)\.184,54)\.186,29)\.186,60)\.186,61)\.186,65)\.188,45)
\.188,46)\.188,68)\.190,35)\.190,48)\.190,50)\.190,55)\.190,71)
\.192,31)\.192,32)\.192,36)\.192,39)\.192,44)\.192,52)\.192,55)
\.192,56)\.192,57)\.192,61)\.192,63)\.192,64)\.192,67)\.192,68)
\.192,92)\.194,47)\.196,50)\.196,53)\.196,63)\.196,65)\.196,74)
\.196,80)\.198,34)\.198,37)\.198,42)\.198,45)\.198,46)\.198,52)
\.198,53)\.198,59)\.198,64)\.200,49)\.200,51)\.200,53)\.200,55)
\.200,56)\.200,61)\.200,65)\.200,67)\.200,71)\.202,49)\.204,35)
\.204,37)\.204,50)\.204,53)\.204,55)\.204,61)\.204,62)\.204,64)
\.204,65)\.204,100)\.206,50)\.208,47)\.208,52)\.208,53)\.208,56)
\.208,59)\.208,62)\.208,69)\.208,71)\.208,74)\.208,81)\.210,29)
\.210,34)\.210,36)\.210,39)\.210,41)\.210,43)\.210,45)\.210,47)
\.210,49)\.210,53)\.210,55)\.210,59)\.210,64)\.210,65)\.210,67)
\.210,68)\.210,75)\.214,52)\.216,38)\.216,44)\.216,46)\.216,47)
\.216,48)\.216,49)\.216,56)\.216,58)\.216,59)\.216,61)\.216,68)
\.216,73)\.216,75)\.218,53)\.220,45)\.220,47)\.220,53)\.220,56)
\.220,58)\.220,59)\.220,63)\.220,71)\.220,81)\.220,86)\.222,35)
\.222,36)\.222,37)\.222,72)\.222,75)\.224,47)\.224,56)\.224,60)
\.224,61)\.224,66)\.224,68)\.224,75)\.224,76)\.224,80)\.224,90)
\.228,38)\.228,55)\.228,62)\.228,64)\.228,73)\.228,110)\.230,43)
\.230,54)\.230,56)\.232,55)\.232,69)\.232,71)\.234,35)\.234,38)
\.234,44)\.234,50)\.234,53)\.234,56)\.234,59)\.234,63)\.234,65)
\.234,71)\.234,74)\.234,81)\.236,59)\.238,58)\.238,61)\.238,66)
\.238,69)\.240,39)\.240,40)\.240,44)\.240,46)\.240,48)\.240,50)
\.240,51)\.240,52)\.240,53)\.240,54)\.240,56)\.240,61)\.240,62)
\.240,67)\.240,68)\.240,73)\.240,75)\.240,77)\.240,79)\.240,84)
\.240,85)\.240,92)\.242,59)\.242,64)\.244,61)\.246,43)\.246,80)
\.248,74)\.250,51)\.250,61)\.250,65)\.250,67)\.252,43)\.252,44)
\.252,46)\.252,49)\.252,53)\.252,54)\.252,56)\.252,57)\.252,64)
\.252,65)\.252,68)\.252,70)\.252,72)\.252,76)\.252,77)\.252,95)
\.252,101)\.254,62)\.256,68)\.256,80)\.256,84)\.256,90)\.256,92) }{
\.258,42)\.258,43)\.258,44)\.258,62)\.258,63)\.258,64)\.258,85)
\.260,52)\.260,59)\.260,70)\.260,73)\.260,77)\.260,79)\.260,84)
\.260,103)\.262,64)\.264,46)\.264,57)\.264,58)\.264,62)\.264,74)
\.264,76)\.264,79)\.264,82)\.264,83)\.264,92)\.266,56)\.266,77)
\.270,46)\.270,48)\.270,53)\.270,55)\.270,56)\.270,59)\.270,62)
\.270,65)\.270,67)\.270,71)\.270,74)\.270,86)\.272,75)\.272,87)
\.272,89)\.272,92)\.272,97)\.274,67)\.276,43)\.276,99)\.276,131)
\.278,68)\.280,67)\.280,73)\.280,75)\.280,76)\.280,77)\.280,81)
\.280,86)\.280,88)\.280,90)\.282,45)\.282,68)\.282,92)\.284,69)
\.286,70)\.286,71)\.286,76)\.288,53)\.288,62)\.288,68)\.288,77)
\.288,81)\.288,85)\.288,86)\.288,95)\.290,57)\.290,69)\.290,71)
\.290,83)\.290,85)\.294,47)\.294,55)\.294,63)\.294,80)\.294,90)
\.296,91)\.300,53)\.300,55)\.300,60)\.300,65)\.300,74)\.300,80)
\.300,84)\.300,85)\.300,91)\.300,92)\.300,94)\.300,104)\.304,98)
\.304,102)\.304,108)\.306,47)\.306,50)\.306,53)\.306,63)\.306,66)
\.306,69)\.306,79)\.308,80)\.308,81)\.308,105)\.310,76)\.310,93)
\.312,52)\.312,55)\.312,62)\.312,66)\.312,67)\.312,69)\.312,81)
\.312,83)\.312,99)\.314,77)\.316,79)\.316,116)\.318,51)\.318,78)
\.318,103)\.320,69)\.320,84)\.320,90)\.320,93)\.320,101)\.320,108)
\.322,68)\.322,76)\.322,79)\.322,90)\.324,90)\.328,85)\.328,101)
\.328,103)\.330,45)\.330,47)\.330,56)\.330,57)\.330,59)\.330,63)
\.330,65)\.330,67)\.330,69)\.330,71)\.330,86)\.330,93)\.330,109)
\.334,82)\.336,47)\.336,55)\.336,61)\.336,63)\.336,65)\.336,68)
\.336,69)\.336,99)\.336,101)\.336,109)\.338,77)\.338,83)\.340,70)
\.340,88)\.342,56)\.342,71)\.342,74)\.342,77)\.342,108)\.342,110)
\.348,55)\.348,88)\.350,69)\.350,71)\.350,73)\.350,84)\.350,86)
\.350,88)\.350,92)\.350,94)\.352,79)\.352,92)\.352,97)\.354,57)
\.354,59)\.354,86)\.360,61)\.360,62)\.360,65)\.360,74)\.360,76)
\.360,77)\.360,81)\.360,85)\.360,98)\.360,103)\.360,104)\.360,110)
\.360,111)\.362,89)\.364,89)\.364,95)\.366,61)\.366,92)\.368,126)
\.374,79)\.374,87)\.374,97)\.378,53)\.378,65)\.378,66)\.378,68)
\.378,86)\.378,95)\.378,101)\.380,114)\.382,94)\.384,63)\.384,68)
\.384,80)\.384,84)\.384,92)\.390,51)\.390,69)\.390,71)\.390,77)
\.390,78)\.390,79)\.390,103)\.392,106)\.392,125)\.394,97)\.396,65)
\.396,68)\.396,89)\.396,101)\.396,105)\.396,108)\.396,122)\.400,84)
\.400,110)\.400,129)\.402,66)\.406,117)\.408,70)\.408,84)\.408,85)
\.408,86)\.408,89)\.408,105)\.408,121)\.410,103)\.414,65)\.414,68)
\.414,92)\.414,102)\.414,131)\.416,108)\.416,115)\.418,98)\.418,108)
\.418,117)\.420,59)\.420,61)\.420,67)\.420,74)\.420,75)\.420,76)
\.420,78)\.420,82)\.420,90)\.420,91)\.420,122)\.420,131)\.422,104)
\.424,103)\.424,109)\.426,104)\.430,87)\.432,80)\.432,87)\.432,89)
\.432,140)\.442,95)\.442,115)\.448,140)\.450,78)\.450,80)\.450,89)
\.450,99)\.450,104)\.456,95)\.456,98)\.456,143)\.462,65)\.462,77)
\.462,79)\.462,81)\.462,85)\.462,109)\.462,122)\.468,82)\.468,128)
\.470,93)\.474,78)\.474,79)\.474,116)\.478,118)\.480,84)\.480,88)
\.480,92)\.480,94)\.480,101)\.480,105)\.480,109)\.480,129)\.482,119)
\.486,80)\.486,90)\.486,106)\.486,122)\.490,95)\.490,106)\.490,125)
\.494,128)\.496,164)\.498,81)\.498,82)\.504,85)\.504,107)\.506,141)
\.510,84)\.510,86)\.510,92)\.510,95)\.510,99)\.510,109)\.510,121)
\.516,85)\.520,163)\.522,88)\.522,126)\.528,79)\.528,92)\.528,94)
\.528,100)\.528,108)\.528,138)\.530,103)\.530,105)\.530,129)\.534,87)
\.534,88)\.540,98)\.540,114)\.544,140)\.546,83)\.546,85)\.546,95)
\.546,96)\.546,97)\.546,109)\.552,118)\.570,94)\.570,95)\.570,104)
\.570,107)\.570,117)\.570,122)\.574,119)\.574,142)\.576,95)\.576,101)
\.588,106)\.590,115)\.594,89)\.594,98)\.594,101)\.594,138)\.600,84)
\.600,107)\.600,110)\.600,119)\.600,124)\.600,129)\.604,151)\.606,99)
\.606,100)\.618,103)\.630,87)\.630,95)\.630,114)\.630,131)\.630,135)
\.630,137)\.636,103)\.636,109)\.642,105)\.648,134)\.648,196)\.650,143)
\.650,163)\.654,108)\.666,110)\.672,95)\.672,118)\.672,140)\.678,111)
\.682,164)\.690,135)\.696,118)\.696,144)\.714,107)\.714,117)\.714,118)
\.720,138)\.726,119)\.726,129)\.738,122)\.744,158)\.750,103)\.750,119)
\.750,123)\.750,126)\.750,129)\.780,174)\.786,130)\.798,115)\.810,138)
\.810,174)\.810,196)\.816,140)\.822,135)\.834,137)\.834,138)\.840,148)
\.840,158)\.840,176)\.858,151)\.870,147)\.870,158)\.870,167)\.890,175)
\.906,151)\.924,137)\.930,158)\.930,174)\.930,206)\.936,196)\.966,137)
\.966,143)\.966,153)\.966,159)\.966,165)\.1050,149)\.1050,151)\.1050,181)
\.1128,229)\.1170,196)\.1218,173)\.1254,207)\.1302,185)\.1410,229)\.1806,263)
} \ve

\figd  \he=2000 \hopl{
\.5,1)\.6,1)\.7,2)\.8,2)\.9,2)\.9,4)\.9,8)
\.10,3)\.11,4)\.12,2)\.12,3)\.12,4)\.12,5)\.12,9)
\.12,11)\.12,23)\.13,5)\.14,5)\.15,3)\.15,5)\.15,7)
\.15,8)\.15,13)\.15,15)\.15,16)\.15,17)\.15,19)\.15,23)
\.16,5)\.16,8)\.17,7)\.18,5)\.18,7)\.18,9)\.18,11)
\.18,17)\.19,8)\.20,6)\.20,7)\.20,9)\.20,13)\.20,15)
\.20,19)\.21,5)\.21,8)\.21,13)\.21,15)\.21,16)\.21,19)
\.21,21)\.21,23)\.21,24)\.21,25)\.21,26)\.22,9)\.23,10)
\.24,6)\.24,7)\.24,8)\.24,9)\.24,10)\.24,11)\.24,12)
\.24,13)\.24,14)\.24,16)\.24,19)\.24,20)\.24,23)\.24,24)
\.24,26)\.25,9)\.25,11)\.25,13)\.25,25)\.26,11)\.27,8)
\.27,10)\.27,14)\.27,16)\.27,20)\.27,21)\.27,23)\.27,26)
\.27,32)\.28,8)\.28,14)\.28,19)\.28,25)\.29,13)\.30,9)
\.30,11)\.30,13)\.30,15)\.30,17)\.30,19)\.30,23)\.30,32)
\.31,14)\.32,11)\.32,12)\.32,14)\.32,17)\.32,20)\.32,22)
\.33,19)\.33,23)\.33,39)\.34,15)\.35,13)\.35,14)\.35,16)
\.35,17)\.36,9)\.36,10)\.36,11)\.36,12)\.36,13)\.36,14)
\.36,15)\.36,16)\.36,17)\.36,19)\.36,20)\.36,23)\.36,25)
\.36,26)\.36,27)\.36,35)\.36,36)\.37,17)\.39,11)\.39,17)
\.39,20)\.39,23)\.39,27)\.40,13)\.40,15)\.40,16)\.40,17)
\.40,19)\.40,22)\.40,23)\.40,24)\.40,31)\.40,32)\.40,33)
\.41,19)\.42,15)\.42,16)\.42,19)\.42,21)\.42,23)\.42,25)
\.42,26)\.44,19)\.44,20)\.44,35)\.45,11)\.45,15)\.45,17)
\.45,19)\.45,20)\.45,22)\.45,25)\.45,28)\.45,29)\.45,34)
\.48,12)\.48,13)\.48,14)\.48,15)\.48,16)\.48,17)\.48,18)
\.48,20)\.48,21)\.48,22)\.48,23)\.48,24)\.48,26)\.48,27)
\.48,29)\.48,31)\.48,34)\.48,40)\.49,20)\.50,23)\.51,17)
\.51,23)\.51,24)\.52,41)\.54,18)\.54,25)\.54,31)\.56,20)
\.56,22)\.56,24)\.56,28)\.56,29)\.56,32)\.56,43)\.57,20)
\.57,23)\.60,15)\.60,16)\.60,17)\.60,18)\.60,19)\.60,21)
\.60,22)\.60,23)\.60,24)\.60,26)\.60,27)\.60,30)\.60,31)
\.60,34)\.60,35)\.60,37)\.60,38)\.60,41)\.60,44)\.60,51)
\.61,29)\.63,17)\.63,20)\.63,23)\.63,29)\.63,38)\.63,45)
\.64,25)\.64,27)\.64,29)\.64,35)\.64,40)\.65,29)\.66,21)
\.66,35)\.66,41)\.66,44)\.68,51)\.69,43)\.69,44)\.70,33)
\.70,37)\.72,19)\.72,22)\.72,24)\.72,25)\.72,27)\.72,28)
\.72,29)\.72,32)\.72,34)\.72,35)\.72,38)\.72,39)\.72,40)
\.72,41)\.72,44)\.73,35)\.75,23)\.75,24)\.75,27)\.75,31)
\.75,33)\.75,39)\.75,49)\.75,55)\.76,56)\.78,25)\.78,41)
\.78,49)\.80,29)\.80,33)\.80,34)\.80,37)\.80,39)\.80,41)
\.81,44)\.81,46)\.81,56)\.83,40)\.84,17)\.84,23)\.84,27)
\.84,29)\.84,32)\.84,33)\.84,37)\.84,41)\.84,43)\.84,45)
\.84,51)\.84,62)\.85,47)\.85,49)\.87,28)\.87,49)\.87,55)
\.87,56)\.88,41)\.89,43)\.90,27)\.90,29)\.90,31)\.90,43)
\.90,49)\.90,53)\.90,55)\.90,63)\.91,41)\.91,44)\.91,47)
\.96,26)\.96,28)\.96,32)\.96,35)\.96,36)\.96,38)\.96,39)
\.96,40)\.96,52)\.96,62)\.97,47)\.99,37)\.99,46)\.99,53)
\.100,49)\.101,49)\.102,51)\.103,50)\.104,51)\.105,39)\.105,43)
\.105,64)\.105,67)\.105,75)\.107,52)\.108,43)\.108,46)\.108,47)
\.108,56)\.108,61)\.108,63)\.108,65)\.110,57)\.111,37)\.112,42)
\.112,62)\.113,55)\.114,37)\.114,56)\.115,54)\.116,49)\.117,59)
\.117,81)\.120,31)\.120,34)\.120,36)\.120,37)\.120,41)\.120,42)
\.120,43)\.120,48)\.120,49)\.120,55)\.120,57)\.120,59)\.120,60)
\.120,64)\.120,70)\.120,72)\.121,59)\.121,64)\.124,44)\.126,45)
\.126,67)\.128,49)\.128,52)\.128,53)\.128,62)\.129,63)\.131,64)
\.132,41)\.132,42)\.132,46)\.132,54)\.132,62)\.135,46)\.135,53)
\.135,56)\.138,45)\.139,68)\.140,69)\.141,92)\.143,71)\.144,41)
\.144,44)\.144,46)\.144,51)\.144,62)\.144,63)\.144,65)\.144,68)
\.144,74)\.144,86)\.145,71)\.145,83)\.145,85)\.147,63)\.152,66)
\.152,68)\.156,67)\.156,81)\.159,103)\.160,67)\.160,69)\.161,79)
\.161,90)\.163,80)\.165,56)\.165,63)\.165,65)\.168,46)\.168,55)
\.168,58)\.168,62)\.168,70)\.168,75)\.168,90)\.169,83)\.171,56)
\.173,85)\.176,73)\.176,76)\.177,59)\.180,46)\.180,56)\.180,61)
\.180,62)\.180,63)\.180,65)\.180,69)\.180,73)\.180,83)\.180,110)
\.184,76)\.187,97)\.192,52)\.192,54)\.192,62)\.192,66)\.192,73)
\.192,77)\.192,79)\.192,89)\.195,69)\.195,79)\.198,70)\.198,76)
\.200,73)\.204,53)\.204,69)\.204,93)\.207,68)\.208,81)\.210,72)
\.210,80)\.210,94)\.216,64)\.216,68)\.216,77)\.216,86)\.216,101)
\.220,102)\.231,77)\.234,83)\.237,79)\.240,84)\.240,87)\.240,89)
\.240,104)\.241,119)\.243,80)\.243,90)\.252,86)\.252,92)\.255,84)
\.264,85)\.264,89)\.264,126)\.270,91)\.272,115)\.273,95)\.280,106)
\.300,105)\.303,100)\.312,97)\.312,104)\.324,90)\.330,111)\.330,122)
\.336,102)\.342,128)\.352,164)\.360,106)\.360,125)\.360,126)\.375,129)
\.384,110)\.384,122)\.408,118)\.420,134)\.423,140)\.432,138)\.453,151)
\.456,130)\.480,158)\.504,148)\.600,181)\.600,196)} \ve

\fige  \hocucut{
\.1,0)\.1,1)\.2,1)\.2,2)\.2,3)\.2,4)\.2,5)
\.2,6)\.2,7)\.2,9)\.2,10)\.2,11)\.2,23)\.3,2)
\.3,3)\.3,4)\.3,5)\.3,6)\.3,7)\.3,8)\.3,9)
\.3,10)\.3,11)\.3,17)\.4,3)\.4,4)\.4,5)\.4,6)
\.4,7)\.4,8)\.4,9)\.4,10)\.4,11)\.4,12)\.4,13)
\.4,14)\.4,15)\.4,16)\.4,17)\.4,18)\.4,19)\.4,20)
\.4,21)\.4,23)\.4,24)\.4,25)\.4,26)\.4,27)\.4,32)
\.5,4)\.5,9)\.6,5)\.6,6)\.6,7)\.6,8)\.6,9)
\.6,10)\.6,11)\.6,12)\.6,13)\.6,14)\.6,15)\.6,16)
\.6,17)\.6,18)\.6,19)\.6,20)\.6,21)\.6,22)\.6,23)
\.6,24)\.6,25)\.6,26)\.6,27)\.6,29)\.6,31)\.6,33)
\.6,35)\.6,36)\.8,7)\.8,8)\.8,9)\.8,10)\.8,11)
\.8,12)\.8,13)\.8,14)\.8,15)\.8,16)\.8,17)\.8,18)
\.8,19)\.8,20)\.8,21)\.8,22)\.8,23)\.8,24)\.8,25)
\.8,26)\.8,27)\.8,28)\.8,29)\.8,30)\.8,31)\.8,32)
\.8,33)\.8,34)\.8,35)\.8,36)\.8,37)\.8,38)\.8,39)
\.8,40)\.8,41)\.8,44)\.8,51)\.9,8)\.9,9)\.9,10)
\.9,11)\.9,12)\.9,13)\.9,14)\.9,15)\.9,16)\.9,17)
\.9,18)\.9,19)\.9,20)\.9,21)\.9,22)\.9,23)\.9,24)
\.9,25)\.9,26)\.9,27)\.9,31)\.9,32)\.9,39)\.9,41)
\.10,9)\.10,10)\.10,11)\.10,13)\.10,14)\.10,15)\.10,16)
\.10,17)\.10,19)\.10,20)\.10,21)\.10,23)\.10,24)\.10,25)
\.10,26)\.10,27)\.10,29)\.10,30)\.10,31)\.10,34)\.10,35)
\.10,39)\.10,41)\.10,44)\.10,49)\.11,10)\.11,21)\.12,11)
\.12,12)\.12,13)\.12,14)\.12,15)\.12,16)\.12,17)\.12,18)
\.12,19)\.12,20)\.12,21)\.12,22)\.12,23)\.12,24)\.12,25)
\.12,26)\.12,27)\.12,28)\.12,29)\.12,30)\.12,31)\.12,32)
\.12,33)\.12,34)\.12,35)\.12,36)\.12,37)\.12,38)\.12,39)
\.12,40)\.12,41)\.12,42)\.12,43)\.12,44)\.12,45)\.12,46)
\.12,47)\.12,48)\.12,49)\.12,50)\.12,51)\.12,53)\.12,55)
\.12,62)\.12,63)\.12,65)\.14,13)\.14,27)\.15,14)\.15,29)
\.16,15)\.16,16)\.16,17)\.16,18)\.16,19)\.16,20)\.16,21)
\.16,22)\.16,23)\.16,24)\.16,25)\.16,26)\.16,27)\.16,28)
\.16,29)\.16,30)\.16,31)\.16,32)\.16,33)\.16,34)\.16,35)
\.16,36)\.16,37)\.16,38)\.16,39)\.16,40)\.16,41)\.16,42)
\.16,43)\.16,44)\.16,45)\.16,46)\.16,47)\.16,48)\.16,49)
\.16,50)\.16,51)\.16,52)\.16,53)\.16,54)\.16,55)\.16,56)
\.16,57)\.16,58)\.16,59)\.16,60)\.16,62)\.16,64)\.16,65)
\.16,66)\.16,67)\.16,69)\.16,70)\.16,72)\.16,75)\.18,17)
\.18,18)\.18,19)\.18,20)\.18,21)\.18,22)\.18,23)\.18,24)
\.18,25)\.18,26)\.18,27)\.18,28)\.18,29)\.18,30)\.18,31)
\.18,32)\.18,33)\.18,34)\.18,35)\.18,36)\.18,37)\.18,38)
\.18,39)\.18,40)\.18,41)\.18,42)\.18,43)\.18,44)\.18,45)
\.18,46)\.18,47)\.18,48)\.18,49)\.18,50)\.18,52)\.18,53)
\.18,54)\.18,55)\.18,56)\.18,57)\.18,58)\.18,59)\.18,61)
\.18,63)\.18,65)\.18,67)\.18,76)\.20,19)\.20,20)\.20,21)
\.20,22)\.20,23)\.20,24)\.20,25)\.20,26)\.20,27)\.20,28)
\.20,29)\.20,30)\.20,31)\.20,32)\.20,33)\.20,34)\.20,35)
\.20,36)\.20,37)\.20,38)\.20,39)\.20,40)\.20,41)\.20,42)
\.20,43)\.20,44)\.20,45)\.20,46)\.20,47)\.20,48)\.20,49)
\.20,50)\.20,51)\.20,52)\.20,53)\.20,54)\.20,55)\.20,56)
\.20,57)\.20,59)\.20,60)\.20,61)\.20,62)\.20,63)\.20,65)
\.20,66)\.20,67)\.20,69)\.20,74)\.21,20)\.21,23)\.21,26)
\.21,29)\.21,38)\.21,41)\.21,44)\.21,47)\.22,21)\.22,22)
\.22,23)\.22,25)\.22,32)\.22,33)\.22,34)\.22,36)\.22,43)
\.22,44)\.22,45)\.22,54)\.22,56)\.22,89)\.23,22)\.23,45)
\.24,23)\.24,25)\.24,26)\.24,27)\.24,28)\.24,29)\.24,30)
\.24,31)\.24,32)\.24,33)\.24,34)\.24,35)\.24,36)\.24,37)
\.24,38)\.24,39)\.24,40)\.24,41)\.24,42)\.24,43)\.24,44)
\.24,45)\.24,46)\.24,47)\.24,48)\.24,49)\.24,50)\.24,51)
\.24,52)\.24,53)\.24,54)\.24,55)\.24,56)\.24,57)\.24,58)
\.24,59)\.24,60)\.24,61)\.24,62)\.24,63)\.24,64)\.24,65)
\.24,66)\.24,67)\.24,68)\.24,69)\.24,70)\.24,71)\.24,72)
\.24,73)\.24,74)\.24,75)\.24,76)\.24,77)\.24,78)\.24,79)
\.24,80)\.24,81)\.24,82)\.24,83)\.24,84)\.24,85)\.24,86)
\.24,87)\.24,90)\.24,92)\.24,94)\.24,96)\.24,97)\.24,98)
\.24,99)\.24,101)\.24,103)\.24,110)\.26,25)\.26,51)\.27,26)
\.27,27)\.27,28)\.27,29)\.27,32)\.27,33)\.27,34)\.27,35)
\.27,36)\.27,37)\.27,38)\.27,39)\.27,41)\.27,43)\.27,44)
\.27,46)\.27,50)\.27,51)\.27,53)\.27,56)\.27,61)\.27,66)
\.27,67)\.27,69)\.28,27)\.28,28)\.28,31)\.28,34)\.28,41)
\.28,42)\.28,43)\.28,45)\.28,46)\.28,48)\.28,49)\.28,55)
\.28,56)\.28,57)\.28,59)\.28,62)\.28,69)\.28,71)\.29,28)
\.30,29)\.30,30)\.30,32)\.30,33)\.30,34)\.30,35)\.30,37)
\.30,38)\.30,39)\.30,40)\.30,42)\.30,43)\.30,44)\.30,45)
\.30,46)\.30,47)\.30,48)\.30,49)\.30,50)\.30,52)\.30,53)
\.30,54)\.30,59)\.30,60)\.30,61)\.30,63)\.30,64)\.30,65)
\.30,67)\.30,69)\.30,70)\.30,75)\.30,76)\.30,78)\.30,79)
\.30,81)\.30,82)\.30,85)\.30,86)\.30,87)\.30,90)\.30,91)
\.30,92)\.30,93)\.32,31)\.32,32)\.32,33)\.32,35)\.32,36)
\.32,37)\.32,39)\.32,40)\.32,41)\.32,43)\.32,44)\.32,45)
\.32,46)\.32,47)\.32,48)\.32,49)\.32,50)\.32,51)\.32,52)
\.32,53)\.32,54)\.32,55)\.32,56)\.32,57)\.32,58)\.32,59)
\.32,60)\.32,61)\.32,62)\.32,63)\.32,64)\.32,65)\.32,66)
\.32,67)\.32,68)\.32,69)\.32,70)\.32,71)\.32,72)\.32,73)
\.32,74)\.32,75)\.32,76)\.32,77)\.32,79)\.32,80)\.32,81)
\.32,82)\.32,83)\.32,84)\.32,85)\.32,86)\.32,87)\.32,88)
\.32,89)\.32,90)\.32,91)\.32,92)\.32,93)\.32,96)\.32,97)
\.32,99)\.32,100)\.32,101)\.32,104)\.32,107)\.32,109)\.32,110)
\.32,112)\.32,115)\.32,117)\.32,118)\.32,120)\.32,126)\.32,127)
\.32,135)\.33,32)\.33,65)\.35,34)\.35,69)\.36,35)\.36,36)
\.36,37)\.36,38)\.36,39)\.36,41)\.36,43)\.36,44)\.36,45)
\.36,46)\.36,47)\.36,48)\.36,49)\.36,50)\.36,51)\.36,52)
\.36,53)\.36,54)\.36,55)\.36,56)\.36,57)\.36,58)\.36,59)
\.36,60)\.36,61)\.36,62)\.36,63)\.36,64)\.36,65)\.36,66)
\.36,67)\.36,68)\.36,69)\.36,70)\.36,71)\.36,72)\.36,73)
\.36,74)\.36,75)\.36,76)\.36,77)\.36,79)\.36,80)\.36,81)
\.36,83)\.36,86)\.36,87)\.36,89)\.36,91)\.36,92)\.36,93)
\.36,94)\.36,95)\.36,97)\.36,98)\.36,99)\.36,100)\.36,101)
\.36,102)\.36,103)\.36,104)\.36,105)\.36,106)\.36,107)\.36,108)
\.36,109)\.36,110)\.36,111)\.36,112)\.36,114)\.36,116)\.36,118)
\.36,121)\.36,122)\.36,126)\.36,127)\.36,128)\.39,38)\.40,39)
\.40,40)\.40,43)\.40,45)\.40,46)\.40,47)\.40,49)\.40,51)
\.40,53)\.40,54)\.40,55)\.40,56)\.40,57)\.40,58)\.40,59)
\.40,60)\.40,61)\.40,62)\.40,63)\.40,64)\.40,65)\.40,66)
\.40,67)\.40,68)\.40,69)\.40,70)\.40,71)\.40,72)\.40,73)
\.40,74)\.40,75)\.40,76)\.40,78)\.40,79)\.40,80)\.40,81)
\.40,82)\.40,83)\.40,84)\.40,85)\.40,86)\.40,87)\.40,88)
\.40,89)\.40,90)\.40,91)\.40,92)\.40,93)\.40,94)\.40,95)
\.40,96)\.40,97)\.40,98)\.40,99)\.40,101)\.40,102)\.40,103)
\.40,104)\.40,105)\.40,107)\.40,109)\.40,111)\.40,113)\.40,114)
\.40,118)\.40,121)\.40,122)\.40,123)\.40,124)\.40,125)\.40,126)
\.40,128)\.40,129)\.40,136)\.40,138)\.40,141)\.40,142)\.40,144)
\.40,146)\.41,40)\.41,81)\.42,41)\.42,42)\.42,43)\.42,44)
\.42,45)\.42,47)\.42,48)\.42,50)\.42,51)\.42,53)\.42,55)
\.42,62)\.42,63)\.42,64)\.42,65)\.42,68)\.42,69)\.42,71)
\.42,72)\.42,74)\.42,76)\.42,77)\.42,78)\.42,80)\.42,81)
\.42,84)\.42,85)\.42,86)\.42,90)\.42,91)\.42,92)\.42,95)
\.42,96)\.42,97)\.42,99)\.42,101)\.42,103)\.42,105)\.42,106)
\.42,122)\.42,126)\.42,127)\.44,43)\.44,45)\.44,47)\.44,54)
\.44,56)\.44,65)\.44,67)\.44,69)\.44,76)\.44,81)\.44,82)
\.44,87)\.44,89)\.44,91)\.44,98)\.44,99)\.44,102)\.44,103)
\.44,109)\.44,130)\.44,131)\.46,45)\.46,46)\.46,68)\.46,69)
\.46,71)\.46,91)\.46,92)\.46,93)\.48,47)\.48,50)\.48,51)
\.48,52)\.48,53)\.48,55)\.48,56)\.48,58)\.48,59)\.48,60)
\.48,61)\.48,62)\.48,63)\.48,64)\.48,65)\.48,66)\.48,67)
\.48,68)\.48,69)\.48,70)\.48,71)\.48,72)\.48,73)\.48,74)
\.48,75)\.48,76)\.48,77)\.48,78)\.48,79)\.48,80)\.48,81)
\.48,82)\.48,83)\.48,84)\.48,85)\.48,86)\.48,87)\.48,88)
\.48,89)\.48,90)\.48,91)\.48,92)\.48,94)\.48,95)\.48,96)
\.48,97)\.48,98)\.48,99)\.48,100)\.48,101)\.48,102)\.48,103)
\.48,104)\.48,105)\.48,106)\.48,107)\.48,108)\.48,109)\.48,110)
\.48,111)\.48,112)\.48,113)\.48,114)\.48,115)\.48,116)\.48,117)
\.48,118)\.48,119)\.48,120)\.48,121)\.48,122)\.48,123)\.48,124)
\.48,125)\.48,126)\.48,127)\.48,129)\.48,130)\.48,131)\.48,133)
\.48,134)\.48,135)\.48,137)\.48,140)\.48,141)\.48,143)\.48,144)
\.48,145)\.48,146)\.48,147)\.48,148)\.48,150)\.48,151)\.48,152)
\.48,153)\.48,154)\.48,155)\.48,156)\.48,157)\.48,158)\.48,161)
\.48,163)\.48,164)\.48,173)\.48,174)\.48,176)\.48,178)\.48,184)
\.48,190)\.48,194)\.48,198)\.48,202)\.48,204)\.48,208)\.50,49)
\.50,51)\.50,59)\.50,61)\.50,63)\.50,65)\.50,67)\.50,73)
\.50,99)\.50,123)\.51,50)\.51,101)\.52,51)\.52,52)\.52,78)
\.52,79)\.52,103)\.52,129)\.52,131)\.53,52)\.54,53)\.54,56)
\.54,62)\.54,65)\.54,66)\.54,68)\.54,70)\.54,71)\.54,72)
\.54,73)\.54,74)\.54,77)\.54,78)\.54,80)\.54,82)\.54,83)
\.54,85)\.54,86)\.54,87)\.54,90)\.54,91)\.54,95)\.54,98)
\.54,99)\.54,100)\.54,101)\.54,103)\.54,104)\.54,107)\.54,108)
\.54,110)\.54,113)\.54,116)\.54,117)\.54,118)\.54,120)\.54,122)
\.54,124)\.54,125)\.54,128)\.54,131)\.54,142)\.54,143)\.54,150)
\.54,152)\.54,154)\.54,155)\.54,157)\.54,173)\.54,180)\.55,59)
\.55,64)\.55,114)\.55,129)\.56,55)\.56,57)\.56,69)\.56,71)
\.56,73)\.56,83)\.56,85)\.56,87)\.56,88)\.56,100)\.56,101)
\.56,111)\.56,115)\.56,117)\.56,125)\.56,129)\.56,147)\.58,57)
\.58,59)\.58,86)\.58,87)\.58,88)\.58,115)\.58,117)\.60,59)
\.60,60)\.60,61)\.60,62)\.60,65)\.60,66)\.60,68)\.60,69)
\.60,70)\.60,71)\.60,72)\.60,73)\.60,74)\.60,75)\.60,76)
\.60,77)\.60,78)\.60,79)\.60,80)\.60,81)\.60,83)\.60,84)
\.60,85)\.60,86)\.60,87)\.60,88)\.60,89)\.60,90)\.60,91)
\.60,92)\.60,93)\.60,94)\.60,95)\.60,96)\.60,97)\.60,98)
\.60,99)\.60,101)\.60,102)\.60,104)\.60,105)\.60,107)\.60,108)    }{
\.60,109)\.60,111)\.60,112)\.60,113)\.60,114)\.60,115)\.60,116)
\.60,117)\.60,118)\.60,120)\.60,121)\.60,122)\.60,123)\.60,125)
\.60,129)\.60,132)\.60,133)\.60,134)\.60,145)\.60,146)\.60,147)
\.60,149)\.60,158)\.60,160)\.60,165)\.60,172)\.60,173)\.60,185)
\.60,186)\.60,190)\.60,218)\.63,62)\.64,63)\.64,67)\.64,68)
\.64,69)\.64,70)\.64,75)\.64,80)\.64,81)\.64,83)\.64,84)
\.64,85)\.64,86)\.64,87)\.64,88)\.64,89)\.64,90)\.64,91)
\.64,92)\.64,93)\.64,94)\.64,95)\.64,97)\.64,99)\.64,100)
\.64,101)\.64,102)\.64,103)\.64,105)\.64,107)\.64,108)\.64,109)
\.64,110)\.64,112)\.64,113)\.64,115)\.64,116)\.64,117)\.64,118)
\.64,119)\.64,120)\.64,121)\.64,122)\.64,123)\.64,124)\.64,126)
\.64,127)\.64,129)\.64,131)\.64,132)\.64,134)\.64,137)\.64,139)
\.64,140)\.64,141)\.64,143)\.64,144)\.64,145)\.64,148)\.64,150)
\.64,151)\.64,152)\.64,153)\.64,155)\.64,158)\.64,161)\.64,165)
\.64,169)\.64,170)\.64,173)\.64,175)\.64,183)\.64,185)\.64,186)
\.64,200)\.64,204)\.64,208)\.64,227)\.64,229)\.65,64)\.66,65)
\.66,66)\.66,67)\.66,68)\.66,71)\.66,76)\.66,79)\.66,82)
\.66,90)\.66,92)\.66,98)\.66,101)\.66,102)\.66,103)\.66,106)
\.66,123)\.66,131)\.66,133)\.66,143)\.66,152)\.66,153)\.66,164)
\.66,165)\.66,183)\.68,67)\.68,135)\.69,68)\.70,69)\.70,70)
\.70,104)\.72,71)\.72,72)\.72,73)\.72,74)\.72,75)\.72,77)
\.72,79)\.72,80)\.72,82)\.72,83)\.72,84)\.72,85)\.72,87)
\.72,89)\.72,91)\.72,94)\.72,95)\.72,96)\.72,97)\.72,98)
\.72,102)\.72,103)\.72,104)\.72,105)\.72,106)\.72,107)\.72,108)
\.72,109)\.72,110)\.72,111)\.72,113)\.72,114)\.72,115)\.72,116)
\.72,117)\.72,118)\.72,119)\.72,121)\.72,122)\.72,123)\.72,124)
\.72,125)\.72,126)\.72,128)\.72,129)\.72,130)\.72,131)\.72,132)
\.72,133)\.72,134)\.72,135)\.72,137)\.72,138)\.72,139)\.72,140)
\.72,141)\.72,142)\.72,143)\.72,144)\.72,145)\.72,147)\.72,148)
\.72,149)\.72,155)\.72,157)\.72,163)\.72,164)\.72,165)\.72,168)
\.72,173)\.72,174)\.72,176)\.72,178)\.72,179)\.72,180)\.72,181)
\.72,182)\.72,185)\.72,186)\.72,187)\.72,190)\.72,196)\.72,200)
\.72,210)\.72,212)\.72,214)\.72,218)\.72,220)\.72,221)\.72,230)
\.72,241)\.72,242)\.72,251)\.72,263)\.72,264)\.72,271)\.74,73)
\.75,74)\.78,77)\.78,78)\.78,79)\.78,81)\.78,83)\.78,89)
\.78,95)\.78,116)\.78,118)\.78,156)\.78,161)\.80,79)\.80,81)
\.80,83)\.80,84)\.80,85)\.80,87)\.80,89)\.80,92)\.80,94)
\.80,95)\.80,97)\.80,100)\.80,101)\.80,103)\.80,105)\.80,107)
\.80,108)\.80,109)\.80,110)\.80,113)\.80,114)\.80,117)\.80,119)
\.80,122)\.80,123)\.80,124)\.80,129)\.80,131)\.80,135)\.80,136)
\.80,138)\.80,141)\.80,144)\.80,145)\.80,147)\.80,151)\.80,154)
\.80,155)\.80,157)\.80,159)\.80,164)\.80,165)\.80,166)\.80,167)
\.80,168)\.80,169)\.80,174)\.80,175)\.80,181)\.80,185)\.80,191)
\.80,196)\.80,198)\.80,214)\.80,217)\.80,231)\.80,238)\.80,243)
\.80,318)\.81,80)\.81,90)\.81,92)\.81,106)\.81,107)\.81,111)
\.81,118)\.81,122)\.81,124)\.81,214)\.82,81)\.82,82)\.82,83)
\.82,124)\.83,82)\.84,83)\.84,85)\.84,86)\.84,87)\.84,88)
\.84,89)\.84,95)\.84,97)\.84,98)\.84,100)\.84,103)\.84,104)
\.84,106)\.84,111)\.84,114)\.84,115)\.84,117)\.84,119)\.84,121)
\.84,124)\.84,125)\.84,126)\.84,127)\.84,128)\.84,149)\.84,152)
\.84,155)\.84,167)\.84,170)\.84,173)\.84,176)\.84,181)\.84,190)
\.84,193)\.86,85)\.88,87)\.88,88)\.88,93)\.88,113)\.88,114)
\.88,116)\.88,118)\.88,126)\.88,129)\.88,130)\.88,131)\.88,132)
\.88,133)\.88,135)\.88,137)\.88,140)\.88,141)\.88,147)\.88,148)
\.88,153)\.88,154)\.88,163)\.88,164)\.88,165)\.88,166)\.88,175)
\.88,184)\.88,185)\.88,243)\.89,88)\.90,89)\.90,94)\.90,98)
\.90,101)\.90,102)\.90,103)\.90,105)\.90,107)\.90,108)\.90,112)
\.90,113)\.90,117)\.90,123)\.90,124)\.90,126)\.90,130)\.90,132)
\.90,135)\.90,137)\.90,138)\.90,144)\.90,164)\.90,185)\.90,207)
\.92,91)\.92,93)\.92,95)\.92,116)\.92,166)\.92,185)\.92,188)
\.92,210)\.95,94)\.96,95)\.96,96)\.96,99)\.96,101)\.96,102)
\.96,103)\.96,107)\.96,108)\.96,109)\.96,111)\.96,115)\.96,117)
\.96,118)\.96,119)\.96,121)\.96,122)\.96,123)\.96,125)\.96,126)
\.96,128)\.96,137)\.96,138)\.96,140)\.96,144)\.96,145)\.96,146)
\.96,148)\.96,149)\.96,151)\.96,153)\.96,155)\.96,157)\.96,158)
\.96,160)\.96,161)\.96,163)\.96,167)\.96,168)\.96,170)\.96,174)
\.96,176)\.96,180)\.96,181)\.96,185)\.96,194)\.96,195)\.96,199)
\.96,200)\.96,201)\.96,206)\.96,208)\.96,212)\.96,214)\.96,218)
\.96,222)\.96,227)\.96,229)\.96,231)\.96,241)\.96,242)\.96,250)
\.96,264)\.96,271)\.96,272)\.96,275)\.96,287)\.96,301)\.96,321)   }{
\.96,330)\.96,355)\.96,376)\.98,97)\.99,98)\.100,99)\.100,100)
\.100,103)\.100,109)\.100,119)\.100,123)\.100,124)\.100,126)\.100,127)
\.100,129)\.100,134)\.100,136)\.100,138)\.100,143)\.100,145)\.100,146)
\.100,148)\.100,149)\.100,150)\.100,151)\.100,161)\.100,184)\.102,102)
\.102,103)\.104,103)\.104,105)\.104,109)\.104,129)\.104,131)\.104,133)
\.104,157)\.104,159)\.104,160)\.105,104)\.106,105)\.108,107)\.108,108)
\.108,110)\.108,113)\.108,115)\.108,116)\.108,122)\.108,128)\.108,131)
\.108,132)\.108,134)\.108,138)\.108,140)\.108,143)\.108,145)\.108,147)
\.108,152)\.108,155)\.108,161)\.108,163)\.108,165)\.108,167)\.108,168)
\.108,172)\.108,173)\.108,174)\.108,175)\.108,176)\.108,190)\.108,194)
\.108,195)\.108,196)\.108,197)\.108,198)\.108,218)\.108,221)\.108,228)
\.108,259)\.108,260)\.108,275)\.108,320)\.110,114)\.110,119)\.110,123)
\.110,125)\.110,129)\.110,130)\.110,131)\.110,136)\.110,141)\.110,165)
\.110,176)\.110,184)\.111,110)\.112,111)\.112,117)\.112,118)\.112,119)
\.112,144)\.112,147)\.112,148)\.112,150)\.112,152)\.112,153)\.112,158)
\.112,159)\.112,167)\.112,173)\.112,195)\.112,204)\.112,205)\.112,210)
\.112,251)\.113,112)\.114,113)\.116,115)\.116,117)\.116,146)\.116,179)
\.119,118)\.120,119)\.120,121)\.120,122)\.120,127)\.120,129)\.120,131)
\.120,132)\.120,137)\.120,138)\.120,140)\.120,142)\.120,143)\.120,145)
\.120,149)\.120,151)\.120,153)\.120,154)\.120,155)\.120,158)\.120,159)
\.120,161)\.120,163)\.120,164)\.120,165)\.120,167)\.120,169)\.120,173)
\.120,174)\.120,176)\.120,177)\.120,180)\.120,181)\.120,184)\.120,186)
\.120,187)\.120,190)\.120,194)\.120,200)\.120,206)\.120,207)\.120,210)
\.120,212)\.120,213)\.120,214)\.120,218)\.120,222)\.120,224)\.120,225)
\.120,229)\.120,230)\.120,231)\.120,236)\.120,246)\.120,257)\.120,301)
\.120,302)\.120,306)\.125,124)\.126,125)\.126,126)\.126,131)\.126,146)
\.126,149)\.126,155)\.126,174)\.126,188)\.126,193)\.126,212)\.126,263)
\.128,127)\.128,136)\.128,140)\.128,141)\.128,151)\.128,169)\.128,176)
\.128,186)\.128,191)\.128,201)\.128,207)\.128,222)\.128,238)\.128,245)
\.128,260)\.128,264)\.130,129)\.130,130)\.131,130)\.132,137)\.132,143)
\.132,153)\.132,154)\.132,159)\.132,165)\.132,166)\.132,183)\.132,212)
\.132,269)\.132,280)\.134,133)\.135,134)\.136,135)\.136,137)\.136,143)
\.136,151)\.138,137)\.138,138)\.138,139)\.138,140)\.138,142)\.138,163)
\.138,166)\.140,141)\.140,144)\.140,153)\.140,167)\.140,172)\.140,212)
\.141,140)\.144,143)\.144,145)\.144,147)\.144,152)\.144,155)\.144,157)
\.144,160)\.144,173)\.144,181)\.144,183)\.144,184)\.144,185)\.144,187)
\.144,188)\.144,194)\.144,196)\.144,197)\.144,198)\.144,202)\.144,212)
\.144,214)\.144,217)\.144,220)\.144,222)\.144,227)\.144,229)\.144,230)
\.144,236)\.144,239)\.144,245)\.144,250)\.144,263)\.144,264)\.144,266)
\.144,271)\.144,275)\.144,291)\.144,311)\.144,335)\.146,145)\.147,149)
\.147,176)\.148,148)\.150,151)\.150,159)\.150,164)\.150,174)\.150,179)
\.153,152)\.155,154)\.156,155)\.156,157)\.156,158)\.156,167)\.156,168)
\.156,173)\.156,174)\.156,175)\.156,180)\.156,184)\.156,196)\.156,200)
\.156,275)\.156,330)\.158,157)\.160,175)\.160,196)\.160,218)\.160,219)
\.160,229)\.160,246)\.160,259)\.160,291)\.160,301)\.162,162)\.162,163)
\.162,165)\.162,178)\.162,191)\.162,228)\.162,245)\.162,275)\.164,165)
\.164,204)\.164,206)\.165,164)\.166,166)\.168,167)\.168,173)\.168,174)
\.168,176)\.168,179)\.168,181)\.168,195)\.168,208)\.168,230)\.168,254)
\.168,348)\.171,179)\.174,173)\.174,176)\.174,179)\.176,175)\.176,177)
\.176,183)\.176,186)\.176,194)\.178,178)\.180,180)\.180,182)\.180,185)
\.180,188)\.180,194)\.180,202)\.180,206)\.180,207)\.180,208)\.180,214)
\.180,218)\.180,222)\.180,228)\.180,231)\.180,236)\.180,246)\.180,280)
\.180,282)\.180,291)\.180,302)\.180,330)\.184,229)\.184,234)\.184,321)
\.186,185)\.189,188)\.191,190)\.192,192)\.192,194)\.192,205)\.192,214)
\.192,254)\.192,259)\.192,260)\.192,271)\.192,282)\.192,299)\.192,330)
\.192,377)\.194,193)\.196,196)\.198,198)\.198,200)\.200,199)\.200,201)
\.200,229)\.200,301)\.208,264)\.208,265)\.210,210)\.210,212)\.210,214)
\.210,230)\.210,251)\.212,213)\.216,218)\.216,222)\.216,224)\.216,245)
\.216,252)\.216,254)\.216,258)\.216,269)\.216,275)\.219,218)\.220,246)
\.220,252)\.220,258)\.222,222)\.222,223)\.224,225)\.224,231)\.224,238)
\.224,245)\.224,341)\.226,227)\.228,228)\.230,229)\.232,294)\.232,295)
\.232,318)\.232,376)\.238,238)\.240,240)\.240,243)\.240,259)\.240,263)
\.240,291)\.240,299)\.240,326)\.240,335)\.240,387)\.252,251)\.252,257)
\.252,258)\.252,263)\.252,272)\.252,287)\.252,306)\.252,387)\.256,271)
\.264,266)\.270,270)\.272,272)\.272,275)\.272,291)\.276,287)\.276,321)
\.276,416)\.285,284)\.288,287)\.290,318)\.294,293)\.294,348)\.300,301)
\.300,302)\.306,311)\.312,320)\.330,330)\.336,341)\.348,348)\.348,355)
\.348,376)\.360,377)\.378,377)\.384,387)\.396,416)\.396,462)\.414,416)
\.420,462)\.432,433)\.462,462)\.492,491)}
\end{document}